\documentclass[10pt]{article}

\usepackage{color} 
\usepackage{graphicx}
\usepackage{amsmath}
\usepackage{bbm}
\usepackage{amsfonts}
\usepackage{amssymb}
\usepackage{amsbsy}
\usepackage[T1]{fontenc}
\usepackage{theorem}
\usepackage{stmaryrd}
\usepackage{tipa}
\usepackage{textcomp}
\usepackage{subfigure}
\usepackage{epsfig}
\usepackage[sort,colon]{natbib}
\usepackage{lmodern}
\usepackage{framed}
\usepackage[subfigure]{tocloft}
\usepackage{subeqnarray}
\usepackage{alphalph}
\usepackage[refpage,cfg,prefix]{nomencl}
\usepackage{multirow}
\usepackage{xifthen}
\usepackage{enumerate}
\usepackage{color}
\usepackage{wasysym}
\usepackage{bibentry}
\usepackage{todonotes}
\nobibliography*
\usepackage{verbatim}
\usepackage{booktabs}
\usepackage[hyperindex,breaklinks]{hyperref}
\hypersetup{
colorlinks=true,
urlcolor=blue,
linkcolor=black,
citecolor=black,
}
\usepackage{breakurl}
\usepackage{url}
\usepackage{setspace}
\usepackage{xr}
\makenomenclature

\usepackage[labelfont=bf,labelsep=period,justification=raggedright]{caption}

\makeatletter
\renewcommand{\@biblabel}[1]{\quad#1.}
\makeatother

\date{}

\newcommand{\D}[2]{\frac{\partial #1}{\partial #2}}
\newcommand{\DD}[2]{\frac{\partial^2 #1}{\partial #2^2}}

\newcommand{\ud}{\mbox{d}}

\newcommand{\listofalgorithms}{\textbf{\Huge{List of Algorithms}}}
\newlistof{Algorithm}{exp}{\listofalgorithms}
\newcounter{instructioncounter}

\begin{document}

\begin{flushleft}
{\Large
\textbf{The pseudo-compartment method for coupling PDE and compartment-based models of diffusion}
}
\\
Christian A. Yates$^{1,\ast}$, 
Mark B. Flegg$^{2}$
\\
\bf{1} Department of Mathematical Sciences, University of Bath, Claverton Down, Bath, BA2 7AY, United Kingdom \\
\bf{2} School of Mathematical Sciences, Monash University, Wellington Road, Clayton, Victoria, 3800, Australia
\\
$\ast$ E-mail: c.yates@bath.ac.uk
\end{flushleft}

Key index words: hybrid modelling, stochastic reaction-diffusion, pseudo-compartment, multiscale modelling

\section*{Abstract}

Spatial reaction-diffusion models have been employed to describe many emergent phenomena in biological systems. The modelling technique most commonly adopted in the literature implements systems of partial differential equations (PDEs), which assumes there are sufficient densities of particles that a continuum approximation is valid. However, due to recent advances in computational power, the simulation, and therefore postulation, of computationally intensive individual-based models has become a popular way to investigate the effects of noise in reaction-diffusion systems in which regions of low copy numbers exist.

The specific stochastic models with which we shall be concerned in this manuscript are referred to as `compartment-based' or `on-lattice'. These models are characterised by a discretisation of the computational domain into a grid/lattice of `compartments'. Within each compartment particles are assumed to be well-mixed and are permitted to react with other particles within their compartment or to transfer between neighbouring compartments. 

Stochastic models provide accuracy but at the cost of significant computational resources. For models which have regions of both low and high concentrations it is often desirable, for reasons of efficiency, to employ coupled multi-scale modelling paradigms. 

In this work we develop two hybrid algorithms in which a PDE in one region of the domain is coupled to a compartment-based model in the other. Rather than attempting to balance average fluxes, our algorithms answer a more fundamental question: `how are individual particles transported between the vastly different model descriptions?' First, we present an algorithm derived by carefully re-defining the continuous PDE concentration as a probability distribution. Whilst this first algorithm shows very strong convergence to analytic solutions of test problems, it can be cumbersome to simulate. Our second algorithm is a simplified and more efficient implementation of the first, it is derived in the continuum limit over the PDE region alone. We test our hybrid methods for functionality and accuracy in a variety of different scenarios by comparing the averaged simulations to analytic solutions of PDEs for mean concentrations.

\section{Introduction}

 Diffusion of macromolecules and larger entities is the result of a rapid number of collisions with solvent molecules. Often, the detailed information about the solvent molecules is not known and the macromolecule (particle) appears to move stochastically. The motion of an individual particle can be simulated two different ways \citep{erban2009smr}; those in which trajectories are defined on continuous domains and those in which trajectories are defined by a random walk on a lattice. Stochastic simulation of individual particles is appropriate on time-scales which are much larger than the time between solvent molecule collisions and, in such circumstances, the particle trajectories may resemble a Wiener process (Brownian motion). Coarse-graining the continuous domain into a lattice of compartments, between which particles undergo a random walk, is a simplification which may be appropriate if the lattice spacing is not too large as to impair good spatial resolution. 
Both off-lattice \citep{andrews2004ssc,erban2009smr,zon2005sbn} and on-lattice \citep{drawert2010dfs,lampoudi2009msa,baker2009fmm,yates2012gfm,yates2013ivd,engblom2009ssr} individual-based stochastic simulations have remained popular paradigms for stochastic diffusion simulations.

It is also common for mathematical models of diffusive processes to treat diffusion using deterministic continuous partial differential equations (PDEs) \citep{murray2003mbs}. That is, it is assumed that the copy numbers of particles are so large that a continuous distribution of particles can be defined which evolves deterministically \citep{berg1993rwb}. The effect on the time derivative of the concentration distribution of particles due to the random motion of individuals is represented by a Laplacian term in the governing equation. Other effects, such as particle flow, reactions and other interactions can be incorporated into this mathematical framework by the introduction of additional terms in the governing PDEs. The simplicity with which diffusion can be represented in these deterministic models has, in part, lead to its popularity in modelling complex physical behaviour. This is notably the case in mathematical modelling of biological processes \citep{field1974ocs,adomian1995dbe,schnakenberg1979scr} 
due to the importance of diffusion of either chemicals and/or cells in many biological environments. 
Some deterministic PDE models have the distinct advantage over stochastic models of diffusion that they are analytically tractable so that the particle dynamics may, in some circumstances, be predicted without numerical simulation \citep{zauderer1983pde}. In addition, when numerical solutions are required, there exists a plethora of well-established techniques at hand \citep{thomas1995npd}.

Whilst PDE interpretations of the diffusion mechanisms are convenient and popular they lack the ability to appropriately represent systems in which stochasticity is inherent \citep{erban2007pgs}. For example, fluctuations in the number of particles in a system are especially important for systems in which one or more of the constituent species is in low abundance \citep{gillespie2013psa}. Such situations appear frequently in biological applications due to the small spatial scales that are often involved and can lead to dynamics which are significantly different from a deterministic model which may be postulated based purely on macroscopic considerations and assuming large copy numbers \citep{swain2002iec,elowitz2002sge,ghosh2007msd,tian2004bsl,vilar2002mnr,gonze2002rcr}. 
The effects of stochasticity have been investigated in spatially extended systems of the type we consider in this paper \citep{baker2009fmm,yates2012gfm,woolley2011psm,gillespie2013psa}. Due to the increase in computer capabilities, stochastic simulation techniques for reaction-diffusion systems have found popularity and have been implemented in a number of freely available simulation packages \citep{ander2004sfs,hattne2005srd,ramsey2005dss,drawert2012urd,sanft2011ssd}. 

Individual-based simulations (off- and on-lattice stochastic simulations) of diffusion are capable of capturing stochastic effects which PDEs cannot. However, this is at the cost of computational effort which becomes an insurmountable barrier very quickly as copy numbers increase \citep{gillespie2013psa}. For extended domains, in which concentrations of particle species vary dramatically, a computationally feasible model may require localised regions in which diffusion is modelled using a continuous PDE (for regions in which concentration is high) and an individual-based simulation (for regions in which concentration is low). For example, in modelling calcium-induced calcium release small numbers of ions bind to and facilitate the opening of ion channels thus inducing the release of numbers of orders of magnitude more calcium ions. An individual approach is appropriate for the initial binding, but infeasible for modelling the concentration of ions after the release \citep{flegg2013dst}. Similar multi-scale 
issues occur in the spatio-temporal modelling of cellular signal transduction \citep{klann2012hsg}, in actin dynamics in filopodia \citep{erban2013msr} and in the formation of morphogen gradients in the early embryo \citep{wolpert1969pis,tostevin2007flp,Kicheva2007kmg,Bollenbach2005rfm}.

The importance of efficient and accurate hybrid algorithms for coupling models of diffusion between different regions of space has resulted in a large body of research in the last decade. Most of the models which couple the PDE paradigm with compartment-based stochastic simulations involve the use of significant overlapping regions and/or the averaging of particle flux in order to simulate the necessary boundary condition for the PDE \citep{flekkoy2001cpf,alexander2002ars,moro2004hms,wagner2004hcf}.  \citet{ferm2010aas} also have a method of coupling a PDE to a compartment-based simulation which employs operator splitting and $\tau$-leaping for the time evolution of the compartment-based model where possible. Recently, hybrid models of stochastic diffusion in which on- and off-lattice models are coupled between different regions of space have been proposed \citep{flegg2012trm,erban2013msr,flegg2013cmc,robinson2014atr,klann2012hsg}. Central to these models is the ability to treat individual particles as they 
migrate from one regime to the other. The individual treatment of particles allows for the extension of these coupling algorithms to low copy number regimes. This type of coupling technique was also recently applied the coupling of a PDE with particles undergoing off-lattice Brownian dynamics \citep{franz2012mrd,franz2013twh}.

In this paper we develop hybrid algorithms which couple a PDE in one region of the domain to a compartment-based model in the other. This allows particle concentrations to be resolved on a fine scale in regions of the domain where an individual-based description is required, but at a much coarser scale in regions where particle numbers are large enough to warrant a continuum description.
Rather than balancing flux at the interface we use a method which is similar to the ghost-cell method \citep{flegg2013cmc}. A defining characteristic of this method is the individual treatment of  particles as they cross the interface. In our coupling method a small region of the PDE domain 
adjacent to the compartment-based region is allowed to contribute particles to the compartment-based regime as if it were itself a compartment. 
 We will refer to this as a pseudo-compartment.
The treatment of this pseudo-compartment is complicated. From the perspective of the PDE region it contains a continuous probability distribution representing the distribution of particles. However, from the perspective of the compartment regime, the pseudo-compartment has compartment-like properties. That is, it contains particles which can jump into their neighbouring compartment according to standard compartment-based rules for diffusion. The pseudo-compartment's duality allows for particles to behave correctly as they cross individually between the two different regimes. When particles cross over the interface a indicator function with the mass of a single particle is added to the probability distribution in the region of the pseudo-compartment. 
 
\begin{figure}[h]
\begin{center} 
\includegraphics[width=\columnwidth]{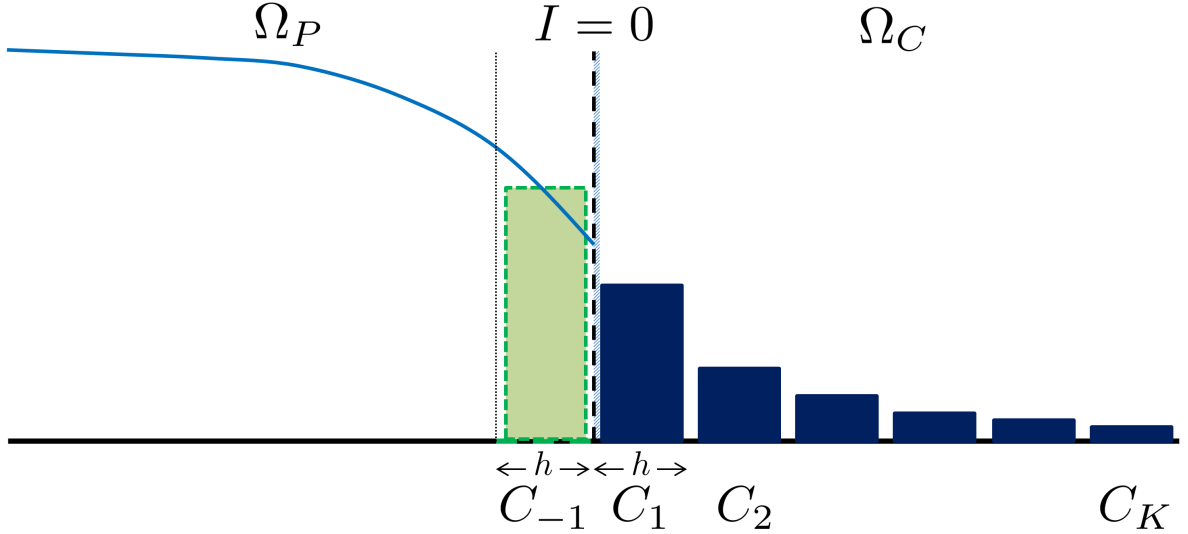}
\end{center}
\caption{In the left-hand-side of the domain, $\Omega_P$, the distribution of particles is modelled as a PDE. On the right-hand-side, $\Omega_C$, the particle density is modelled with a compartment-based approach. The concentration of particles in each compartment, $C_i$ for $i=1,\dots,K$, of $\Omega_C$ is represented by a dark blue rectangle (numbers of particles can be found by multiplying by the compartment width, $h$). The probability density in $\Omega_P$ is represented by the light-blue curve. The two regimes interact through the movement of particles from the left-most compartment in $\Omega_C$, $C_1$, to the pseudo-compartment, $C_{-1}$, (whose average particle concentration is represented by the height of the green box on the right of $\Omega_P$) and vice versa. The height of the green-rectangle is calculated by integrating the PDE solution over the region $[0,h)$ and dividing by $h$. Thus the `number of particles' in the pseudo-compartment can be calculated by multiplying the height of the green box by 
$h$.}
\label{figure:cartoon_schematic_general}
\end{figure}

A one-dimensional graphical representation of this approach is given in Figure \ref{figure:cartoon_schematic_general}. 

We present two separate methods (which we refer to as pseudo-compartment methods (PCM)) for conceptualising and implementing this coupling scheme. The first method is necessary for the precise coupling of systems with low copy numbers. The second method is a mathematically justified simplification of the first for situations in which large particulate copy numbers are present within the PDE domain. We will demonstrate these coupling mechanisms in one dimension only but they are easily generalised to higher dimensional problems with (hyper)-planar interfaces.

\section{Algorithms}

\subsection{Overview of algorithms for deterministic and compartment-based stochastic diffusion}
In this section we will provide an overview of the two different modelling regimes that will be used in our hybrid algorithm, highlighting any intricacies in their definition or implementation.

\subsubsection*{Compartment-based modelling}\label{section:compartment-based_modelling}

We will denote the region in which we employ the compartment-based regime $\Omega_C$. In this region we partition the domain into $K$ regular compartments of length $h$. We choose $h$ to be sufficiently small that particles in each compartment can be considered to be well-mixed, reacting only with particles in their current compartment. We do not keep detailed information about the specific location of each particle, rather we simply store the number of particles of each type in each of these compartments. Whilst this produces a significant saving on computational intensity in comparison to Brownian dynamics models (in which the position of each individual is stored and/or updated at each step), we lose information about the specific location of each of the particles in a compartment and are forced to assume that each particle is distributed uniformly throughout the compartment.

Considering a general system of $N$ species and $M$ chemical reactions, a propensity function, $\alpha_{i,j}(t)$, $i=1,\dots, K$ and $j=1,\dots, M+2$, is assigned to each reaction in each compartment (including the two ``diffusive reactions'' by which particles leave compartment $i$). The propensity functions are defined such that $a_{i,j}(t)dt$ is the probability that the $j^{th}$ reaction occurs in the $i^{th}$ compartment during the infinitesimally small time interval $[t,t + dt]$. The propensity function for a specific reaction in a particular compartment depends on the number of reactant particles available in that compartment (for all but zeroth order reactions), the method by which they combine, the kinetic rate constant and (for all but first order reactions) the size of the compartment, $h$ \citep{erban2009smr}. Similarly, for each species, we can characterise the diffusive jumps of particles between compartment $i$ and its two nearest neighbouring compartments
as a set of first order interactions with propensities 
\begin{equation}\label{jumppropensity}
\alpha_{i,j}=A_iD_{A}/h^2, 
\end{equation} 
for $j=M+1,M+2$. Here, $D_A$ denotes the macroscopic diffusion coefficient of species $A$ and $A_i$ is the copy number of species $A$ in compartment $i$.  Equation (\ref{jumppropensity}) is specific for regular square lattices with no sensing of neighbouring compartments. For unstructured meshes, the jumping propensity $\alpha_{i,j}$ between two lattice points can be calculated by considering a finite element discretisation of the appropriate PDE over the mesh or, in special cases by considering first passage time problems. For details on the derivation of transition rates on unstructured meshes the reader is referred to \citep{engblom2009ssr,yates2013ivd,yates2012gfm}. Such a discretisation justifies the notation $\alpha_{i,j}$ (that is, a dependence on the destination for the jump). In other scenarios, even on regular lattices, (where the particles being modelled are cells, for example) jump rates can depend explicitly on the number of particles in neighbouring compartments in order to mimic ``volume-
filling'' or ``density-dependent'' effects \citep{baker2009fmm}. Although we do not explicitly deal with these scenarios in this work, it would be straightforward to do so.

There are two predominant mechanisms for simulating the time evolution of the system. In the \textit{time-driven} approach, a fixed time-step $\Delta t$ is designated and each reaction $j$ in compartment $i$ occurs with probability $\alpha_{i,j}\Delta t$. This method is simple to implement, but less efficient than the alternative \textit{event-driven} algorithm. In order to ensure accuracy in the time-driven algorithm the time-step must be chosen to be sufficiently small that single particles do not undergo multiple diffusion or reaction events in a single step. This can lead to very small choices of $\Delta t$ and consequently many time-steps in which no reaction/diffusion events occur. In comparison, an event-driven algorithm simulates the evolution of the system by choosing the time-step between reaction/diffusion events from an exponential distribution. The most commonly cited example of such a simulation scheme is the Gillespie's direct method (DM) \citep{gillespie1977ess} in which the time for the next 
reaction of any sort, $\tau$, is chosen from an exponential distribution with parameter $\alpha_0$ (where $\alpha_0=\sum_{i=1}^N\sum_{j=1}^{M+2}\alpha_{i,j}$ is the sum of all the propensity functions). In practice $\tau$ is simulated according to the following formula 
\begin{equation}
 \tau=\frac{1}{\alpha_0}\ln{\frac{1}{r_1}}\label{equation:tau_choice},
\end{equation}
where $r_1$ is uniformly distributed in $(0,1)$. The event that takes place $\tau$ units of time from the current time is chosen from a weighted probability proportional to each of the event propensities $\alpha_{i,j}$. 

There are efficient adaptations of Gillespie's original algorithm available depending on context \citep{cao2004efs,yates2013rrn,mccollum2006sdm,gibson2000ees,li2006ldm}. These algorithms are mathematically identical and differ only in the computational bookkeeping required to implement them. An efficient algorithm designed specifically for spatially extended systems is the next sub-volume method (NSM) from \citet{elf2004ssb}, and an adaptation of the next reaction method (NRM) from \citep{gibson2000ees} (which is itself based on Gillespie's original first reaction method (FRM) \citep{gillespie1976gmn}). For simplicity and efficiency we choose Gillespie's DM (event-driven) in order to evolve particle numbers in $\Omega_C$, although we note that our algorithms can be trivially modified to incorporate the alternative event-driven simulation methods discussed above or time-driven evolution.

\subsubsection*{PDE-based modelling}\label{section:PDE_based_modelling}

We denote the region in which we solve the PDE as $\Omega_P$. In this region we solve the relevant one-dimensional PDE numerically, but to a high degree of accuracy. In particular we solve the PDE
\begin{equation}
\D{p}{t}=D\DD{p}{x}+R(p), \quad \quad x\in \Omega_P,
\end{equation}
where $R(p)$ represents a set of, as yet unspecified, reactions.

A zero-flux boundary condition is implemented on the PDE at the interface between the two regions. This is because flux of particles over this interface is not governed by a continuous diffusive flow of particles but rather a discrete jumping processes which mimics it. It is important to ensure that there is consistency between the methods by which particles transfer back and forth between regimes since any imbalance can create significant numerical artefacts. 

In the examples that follow, we discretise the PDE-domain, $\Omega_P$, into a regular grid with spacing $\Delta x$ where $\Delta x\ll h$. In order that the time-step which we take is not restricted by the requirement for stability of the PDE solver we employ the $\theta$-method with $\theta\leq 1/2$ which is unconditionally stable. In particular we choose $\theta=1/2$  (Crank-Nicolson method) which gives a solution which is second order accurate in space. Note that the coupling algorithm that follows is independent of the method which is used to solve the PDE, providing that method is sufficiently accurate. 

\subsection{Low copy number coupling technique: Algorithm 1}

In both hybrid algorithms presented here, it is important to note that, whilst we implemented the Gillespie direct method for the compartment-based simulation (event-driven) and a Crank-Nicolson method for solving the PDE part of the simulation (both for accuracy reasons), the choice of methodology does not affect the results of the coupling (but may affect the accuracy of the individual simulations).

It is common to think of the PDE description of diffusion as the evolution of a continuous concentration. However, at low concentrations this quantity is not well defined. Here we consider a situation in which there may be insufficient particles to satisfy the continuum limit. According to the definition of concentration, $p(x,t)$, given $N_P$ identical particles in total over the concentration distribution, the probability distribution of finding any particular particle, $p'(x,t)$, is given by $p'(x,t) = p(x,t)/N_P$. The distribution $p'(x,t)$ is well-defined for low copy numbers and behaves in a similar way to the concentration (albeit normalised by $N_P$). In this situation, the continuous distribution which is solved by the PDE cannot be interpreted as a concentration. Instead, we consider $p'(x,t)$ as the probability density to find each of the $N_P$ particles that are within the PDE domain (we denote the number of compartment particles $N_C$ such that total number of particles is $N=N_P+N_C$). We 
specify that in the continuum limit, as $N_P\rightarrow \infty$, the PDE should describe the concentration of particles. As such, we define the continuous distribution $p'(x,t)$ to be the probability density to find each of the $N_P$ PDE-based particles at position $x$, scaled by the number of particles, such that $N_P\int_{\omega\subset\Omega_P} p'(x,t) \ \mathrm{d}x$ is the expected number of particles in an arbitrary subset, $\omega$, of $\Omega_P$.   

Our first algorithm (which is appropriate for situations in which the copy numbers are low) progresses asynchronously. The compartment-based regime is updated in an event-driven manner whilst the PDE domain is updated in a time-driven manner.
We use the model design shown in Figure \ref{figure:cartoon_schematic_general} and describe here how to couple the transition of particles between the PDE and compartment-based regimes using the pseudo-compartment, $C_{-1}$.

Particle numbers in region $\Omega_C$ are updated according to jump propensities given in equation \eqref{jumppropensity}. We extend the definition of these jump propensities to include  transitions to and from the pseudo-compartment, $C_{-1}$, that lies inside the PDE domain. When a particle is chosen to jump left from compartment $C_{1} $ in $\Omega_C$, into the pseudo-compartment, $C_{-1}$, its mass is incorporated into the PDE solution. The expected number of particles within the pseudo-compartment is increased by one by adding $1/h$ to the PDE distribution in $[-h,0)$. 

In $\Omega_C$, the propensity for a jump from compartment to neighbouring compartment is given by equation \eqref{jumppropensity}. We modify the propensity to jump from the pseudo-compartment, $C_{-1}$, into its neighbouring compartment, $C_{1}$, in $\Omega_C$, in order to take account of the expected number of particles in $C_{-1}$. The jump propensity from $C_{-1}$ to $C_{1}$ is therefore given by 
\begin{equation}\label{ghostjump}
\alpha^* = \frac{D}{h^2} A_{-1},
\end{equation} 
where $A_{-1} = \int_{C_{-1}}p(x,t)\ud x$ is the expected number of particles in the pseudo-compartment. For convenience in what follows we denote the normalised number of particles in $C_{-1}$ as $a_{-1}=A_{-1}/N_P$.

Jumps from $C_{-1}$ to $C_{1}$ occur as compartment-based events in the algorithm. This means that in each small time interval $(t,t+dt_p)$ in which the PDE is updated, none of the PDE-based particles can jump into $\Omega_C$. Since $p(x,t)$ is the scaled probability distribution describing the probability of finding these particles in the PDE regime, the new distribution must be conditioned on the event that none of these particles jumped into the compartment-based regime during the PDE update interval $(t,t+dt_p)$. After running the PDE solver on the distribution $p(x,t)$ a change of $\delta_p(x,t)$ must be made to reflect the fact that none of the remaining $N_P$ particles transferred into $C_1$ via $C_{-1}$.

In the small time step $dt_p$ each of the expected $A_{-1}$ particles inside the pseudo-compartment should be capable of jumping over the interface with a probability of $P_{\mathrm{jump}} = Ddt_p/h^2$. By definition 
\begin{align}
\delta_p(x,t) &= N_P\left(Pr(x,t|\mathrm{no \ jump}) - Pr(x,t)\right) \nonumber\\ 
&= \frac{Pr(\mathrm{no \ jump}|x,t)p(x,t)}{Pr(\mathrm{no \ jump})} - p(x,t) \nonumber\\ 
&= \frac{Pr(\mathrm{no \ jump}|x,t)p(x,t)}{1-P_{\mathrm{jump}}a_{-1}+O(dt_p^2)} - p(x,t),
\end{align}
where $Pr$ indicates the probability for a single particle. $Pr(\mathrm{no \ jump}|x,t)$ depends on whether the particle in question is inside or outside the pseudo-compartment, $C_{-1}$: 
\begin{equation}
Pr(\mathrm{no \ jump}|x,t) = 
\begin{cases}
1, & x\notin C_{-1} \\
1-P_{\mathrm{jump}}, & x\in C_{-1}
\end{cases}.
\end{equation}
Therefore,
\begin{align}
\delta_p(x,t) \approx 
\begin{cases}
\dfrac{p(x,t)P_{\mathrm{jump}} a_{-1}}{1-P_{\mathrm{jump}}a_{-1}}, & x\notin C_{-1} \\[1em]
\vspace{0.2cm}
\dfrac{p(x,t)P_{\mathrm{jump}} (a_{-1}-1)}{1-P_{\mathrm{jump}}a_{-1}}, & x\in C_{-1}
\end{cases}.
\end{align}
As $dt_p \rightarrow 0$, $P_{\mathrm{jump}} \rightarrow 0$ and we find that
\begin{align}\label{sinkandsource}
\lim_{\text{d}t_p \rightarrow 0}\frac{\delta_p(x,t)}{dt_p}=
\begin{cases}
\dfrac{D a_{-1}}{h^2}p(x,t), & x\notin C_{-1} \\[1em]
\dfrac{D a_{-1}}{h^2}p(x,t) - \dfrac{D}{h^2}p(x,t), & x\in C_{-1}
\end{cases}.
\end{align}
The PDE is therefore solved with an additional source term of size $p(x,t)D a_{-1}/h^2$, throughout the domain and a sink term inside the pseudo-compartment of size $-p(x,t)D/h^2 $. These additional terms ensure that the probability distribution inside the PDE domain redistributes correctly as a result of not finding pseudo-compartment particles available for jumping in a given time step. 

\begin{figure}[h!!!!!!!!!!!!!!!!!]
\begin{center} 
\includegraphics[width=\columnwidth]{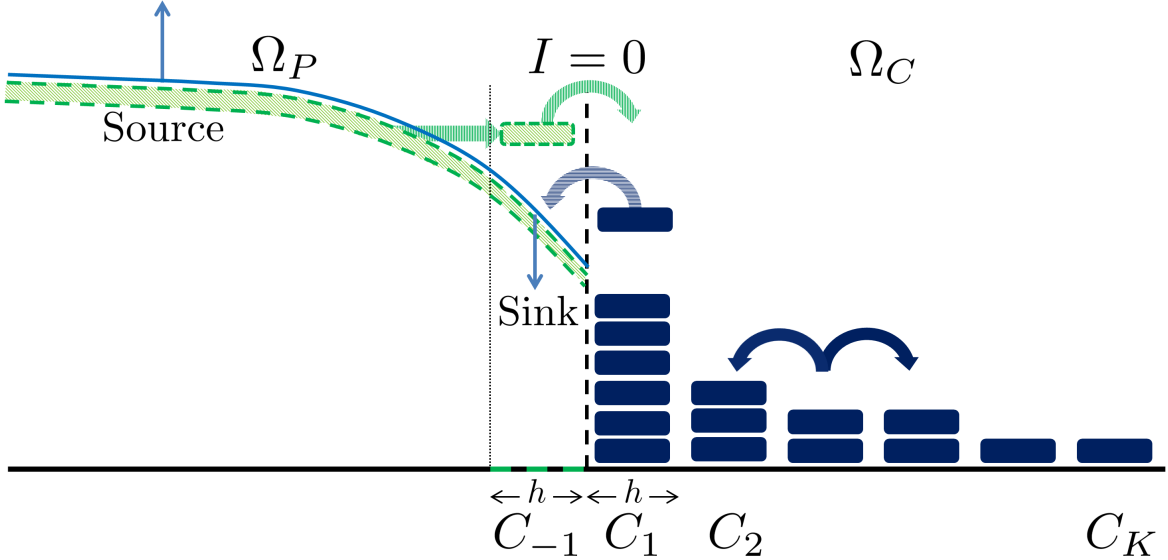}
\end{center}
\caption{A cartoon schematic illustrating the most important features of Algorithm 1 in one dimension. In $\Omega_C$, to the right of the interface, particles are represented as individuals in compartments (dark blue rectangles) with each individual adding $1/h$ to the density in that compartment. Particles can jump to neighbouring compartments within $\Omega_C$ (dark blue block arrows). In $\Omega_P$, to the left of the interface, density is represented by a `continuous' function (light blue curve).  Particles in the left-most compartment in $\Omega_C$ (adjacent to the interface) can jump into the PDE regime (blue horizontally striped arrow) where they are added by means of an indicator function to the PDE solution in the region $[-h,0)$ - the pseudo-compartment. The PDE does not include information about the specific location of individuals and so when a particle jumps from the pseudo-compartment into $\Omega_C$ it must first be realised in the pseudo-compartment by subtraction of its contribution to the 
whole PDE followed by an instantaneous jump over the interface. Every instance in which a particle does not jump from the pseudo-compartment into $\Omega_C$, the conditional probability to find particles in the distribution must necessarily be adjusted so that more weight lies outside of the pseudo-compartment. This effect results in a net source of distribution outside of the pseudo-compartment in the PDE and a balancing net sink of distribution inside the pseudo-compartment such that the total number of particles is conserved in the absence of reactions.}
\label{figure:cartoon_schematic_alg1}
\end{figure}

Changes in the compartment-based regime have an impact on the PDE only when particles cross the interface. We have already discussed the change to the relatively straightforward indicator function addition of density to the PDE when particles cross from $C_1$ to $C_{-1}$. However, when particles cross from $C_{-1}$ to $C_1$ it would be erroneous to simply subtract $1/h$ from the PDE distribution in $C_{-1}$ (reversing the operation for the transfer of particles in the opposing direction). This is because we assume that each of the particles in the PDE region has the same distribution function $p(x,t)/N_P$. Therefore, taking any one of these $N_P$ identical particles across the boundary into $\Omega_C$ will result only in a rescaling of this distribution. Recognising that we lose a single particle's worth of mass, the distribution $p(x,t)$ in $\Omega_P$ should now represent the probability distribution of  $N_P-1$ identical particles so we can calculate the rescaled distribution as
\begin{equation}
 p_{new}(x,t)=\frac{N_P-1}{N_P}p(x,t).\nonumber
\end{equation}
A cartoon schematic illustrating the algorithm is given in Figure \ref{figure:cartoon_schematic_alg1}. The algorithm is also outlined in detail in Table \ref{table:algorithm_1}. It may seem odd that particles appear to be allowed to transferred from the entire PDE domain whilst being allowed only to transfer back into the more localised $C_{-1}$. This is restriction is predicated on the assumption that all particles in the PDE regime have the same distribution which is proportional to the PDE solution, $p(x,t)$. However, when a particle travels from $C_1$ in the compartment-based regime into $C_{-1}$ in the PDE regime, its position is known. This breaks the assumption that all particles in the PDE are identical. This issue has been documented previously \citep{franz2012mrd}. It results in an increase in the variance of the number of molecules near the interface. In \citep{franz2012mrd}, the variance error can be controlled by adding in an `overlap' region in which particles may be in the form of either 
regime (PDE and off-lattice particle-based). The pseudo-compartment in this manuscript plays a similar role to this `overlap' region. In the pseudo-compartment, whilst density is updated using the PDE, particles are transferred by the evolution of the PDE (into the remainder of the PDE regime) and by lattice-based rules (into the compartment-based regime). This complication is unavoidable for algorithms of this nature. The variance is also naturally reduced by increasing the number of particles (see Algorithm 2).

\begin{table}[h!!!!!!!!!!!!!]
\begin{tabular}{p{\textwidth}}
\hline
\begin{enumerate}[(i)]
 \item Initialise the time, $t=t_0$ and set the final time, $t_{final}$. Specify the PDE-update time-step $dt_{P}$ and initialise the next PDE time-step to be $t_P=t+dt_{P}$.
 \item Initialise the number of particles in each compartment in $\Omega_C$, $A_i$ for $i=1,\dots, K$, and the distribution of probability density in $\Omega_P$, $p(x,t_0)$, for $x\in\Omega_P$.
 \item \label{item:calculate_the_propensity_functions_alg1} Calculate the propensity functions for diffusion between the compartment regimes as $\alpha_{i,j}=A_iD/h^2$ for $i=1\dots K$ and $j=M+1, M+2$ and for reactions as $\alpha_{i,j}$ for $i=1\dots K$ and $j=1,\dots, M$ using the usual mass action kinetics.
 \item Calculate the propensity function for diffusion from the pseudo-compartment, $C_{-1}$, in $\Omega_P$ into the adjacent compartment, $C_1$, in $\Omega_C$: $\alpha^*=D	\int_{C_{-1}}p(x,t)\ud x/h^2$ (equation (\ref{ghostjump})).
 \item Calculate the sum of the propensity functions, $\alpha_0=\sum^K_{i=1}\sum^{M+2}_{j=1}\alpha_{i,j}+\alpha^*$.
 \item Determine the time for the next `compartment-based' event $t_C=t+\tau$, where $\tau$ is given by equation \eqref{equation:tau_choice}.
 \item \label{item:update_particle_numbers_in_compartments_alg1} If $t_C<t_P$ then the next compartment-based event occurs:
 \begin{enumerate}[(a)]
 \item Determine which event occurs. Each event occurs with probability proportional to its propensity function (see \citet{gillespie1977ess}).
 \item If the event corresponds to $\alpha_{i,j}$ for $i=1\dots K$ and $j=M+1,M+2$ then move a particle from compartment $i$ in the direction specified by $j$. If the particle crosses the interface into pseudo-compartment $C_{-1}$ then add a particle's worth of mass to the region $C_{-1}$ i.e. $p(x,t+\tau)=p(x,t)+\mathbb{I}_{C_{-1}}/h$. Here $\mathbb{I}_{C_{-1}}$ is the indicator function which takes the value 1 in $C_{-1}$ and 0 elsewhere. 
  \item \label{item:remove_mass_from_PDE_alg1} If the event corresponds to propensity function $\alpha^*$ then place a particle in $C_1$. To remove this particle from the PDE solution, simply rescale the solution i.e. $p(x,t+\tau)=p(x,t)(N_P-1)/N_P $. Update $N_P$ and $N_C$ to reflect the exchange of particles between regimes.
  \item Update the current time, $t=t_C$. 
 \end{enumerate}
\item If $t_P<t_C$ the the PDE regime is updated:
\begin{enumerate}[(a)]
\item Update the PDE solution according to the numerical method described in Section \ref{section:PDE_based_modelling} using $p(x,t)$ as the previous value of the solution. 
\item Add the auxiliary source and sink terms (see equation \eqref{sinkandsource}) to the numerical solution of the PDE.
 \item Update the current time, $t=t_P$ and set the time for the next PDE update step to be $t_P=t_P+dt_P$.
\end{enumerate}
\item If $t\leq t_{final}$, return to step \eqref{item:calculate_the_propensity_functions_alg1}.

Else end.
\end{enumerate}\\
\hline
\end{tabular}
\caption{Algorithm for the coupling of the compartment-based regime with the PDE-based regime in the case of low copy numbers (Algorithm 1).}
\label{table:algorithm_1}
\end{table}

In Algorithm 1, changes to the PDE solution are as a result of diffusion, auxiliary source/sink terms for changes in conditional probability, events where a particle moves from the pseudo-compartment, $C_{-1}$, to the first compartment, $C_1$, in the compartment-based regime, $\Omega_C$, and events where a particle moves in the opposite direction from $C_1$ to $C_{-1}$. All of these changes may be represented formally in the following PDE (in one dimension)
\begin{equation}\label{alg1evolution}
\frac{\partial p(x,t)}{\partial t} = D\frac{\partial^2 p(x,t)}{\partial x^2} +  \left(\frac{N_P^+}{N_P^-}-1\right)p \sum_J \delta(t-t_J) + \frac{Da_{-1}(t)}{h^2}p -  \frac{\mathbb{I}_{C_{-1}}(x)}{h}\left(\frac{D}{h}p - \sum_j \delta(t-t_j)  \right),
\end{equation}
where $\mathbb{I}_{B}(x) = 1$ if $x\in B$ and zero otherwise,   $t_j$ are instants when particles jump from compartment to PDE, $t_J$ are instants when particles jump from PDE to compartment and $N_P^-=N_P(t^-)$ and $N_P^+=N_P(t^+)$ represent numbers of particles in $\Omega_P$ immediately before and after times in which the number discontinuously jumps at respective moments $t_J$ (they can be thought of as left and right limits of $N_p(t)$ respectively). Note that $N_P^+ = N_P^- - 1$. As a common sense check we can integrate over the $\Omega_P$ to derive an equation for the change in the total number of particles represented by the PDE regime, $N_P$:
\begin{equation}
 \frac{\ud N_p}{\ud t}=-\displaystyle\sum_J \delta(t-t_J)+\displaystyle\sum_j \delta(t-t_j).
\end{equation}
As expected, the number of particles changes by integer amounts at the discrete times when particles enter or leave the domain.

\subsection{High copy number coupling technique: Algorithm 2}

The algorithm presented in the previous subsection is accurate for both high and low copy numbers. However, its complexity leads to computational overheads which may prove costly, especially in situations in which copy numbers of species are relatively high and a less sophisticated coupling method would suffice. For this reason we derive the following coupling method (which we refer to as Algorithm 2) as a limiting case of our original algorithm, valid only for the scenario of high copy numbers. 

We predict the benefits of these hybrid methods to be greatest for applications that have regions in which there are low copy numbers (requiring individual-based simulation) and other regions of high copy numbers (in which a PDE description will be more efficient). In particular the copy numbers at the interface should be high in order to justify the use of the PDE there. Arguably if the copy numbers are low at the interface then the interface is positioned incorrectly.

In order to derive our simplified algorithm we allow $N_P \rightarrow \infty$. This implies that times between transfer events (particles crossing from $C_{-1}$ to $C_1$ and vice versa) are low: $0<t_{j+1}-t_{j}\ll 1$ and $0<t_{J+1}-t_{J}\ll 1$. Let us define a small time, $\delta t$, such that 
\begin{enumerate}
\item significantly smaller than the total number of particles in the pseudo-compartment.
\end{enumerate}
Satisfying point 1 ensures that individual jump events may be averaged over $(t,t+\delta t)$ and satisfying point 2 ensures that $\delta t$ is small on the scale of diffusion allowing for PDE solver time discretisation to be implemented at a later stage on the majority of non-jumping particles. Both points 1 and 2 may be satisfied for sufficiently large $N_P$.
 Integrating equation \eqref{alg1evolution} from $t$ to $t+\delta t$, dividing by $\delta t$ and taking the limit as $\delta t\rightarrow 0$ we find
\begin{align}
 \label{intalg1gov}\D{p}{t} &= D\frac{\partial^2 p}{\partial x^2} - \lim_{\delta t\rightarrow 0}\left\{\frac{p}{N_P\delta t} \int_t^{t+\delta t}\sum_J \delta(t'-t_J) \mathrm{d}t' + \frac{Da_{-1}(t)}{h^2}p\right. \\
 & \left.\nonumber \quad \quad \quad \quad \quad \quad -  \frac{D\mathbb{I}_{C_{-1}}(x)}{h^2}p + \lim_{\delta t\rightarrow 0}\frac{\mathbb{I}_{C_{-1}}(x)}{h\delta t}\int_t^{t+\delta t}\sum_j \delta(t'-t_j) \mathrm{d}t'\right\}.
\end{align}
For each $C_{-1}\rightarrow C_{1}$ transfer event we approximate $(N_P^+/N_P^--1)=-1/N_P^-\approx -1/N_P$, since $N_P$ is large. It should be noted that in order to obtain equation (\ref{intalg1gov}) from equation (\ref{alg1evolution}) one requires $\delta t$ be sufficiently small that temporally varying dependant variables are constant on $(t,t+\delta t)$, that is $\int_{t}^{t+\delta t} p(x,t') dt' \approx p(x,t)\delta t$ and similarly for $a_{-1}(t)$ and $1/N_P(t)$. 
 The large number of transfer events that occur between $t$ and $t+\delta t$ mean that the fluctuations in the number of transfer events are negligible and, as such, they can be predicted deterministically using the propensities given in the previous section (equations (\ref{jumppropensity}) and (\ref{ghostjump})):   
 \begin{align}
 \label{simp1}
\int_t^{t+\delta t}\sum_J \delta(t'-t_J) \mathrm{d}t' &= \frac{D\delta t A_{-1}}{h^2}, \quad \quad \mathrm{and} \\
\label{simp2}
 \int_t^{t+\delta t}\sum_j \delta(t'-t_j) \mathrm{d}t' &= \frac{D\delta t A_{1}}{h^2}.
 \end{align}
Substituting equation \eqref{simp1} into equation \eqref{intalg1gov} and recalling that $A_{-1} = a_{-1}N_P$, we find the second and third terms on the right-hand side of equation (\ref{intalg1gov}) cancel each other. Consequently, we see that changes in the PDE as a result of hybridisation with the discrete subdomain (i.e. not including the Laplacian diffusion operator) are confined to within the pseudo-compartment.
  \begin{equation}
 \label{intalg1gov2}\D{p}{t} = D\frac{\partial^2 p(x,t)}{\partial x^2}  -  \lim_{\delta t\rightarrow 0}\left\{\frac{D\mathbb{I}_{C_{-1}}(x)}{h^2}p + \frac{\mathbb{I}_{C_{-1}}(x)}{h\delta t}\int_t^{t+\delta t}\sum_j \delta(t'-t_j) \mathrm{d}t'\right\}.
 \end{equation}
 Substituting equation (\ref{simp1}) into the second term on the right hand side of equation (\ref{intalg1gov2}) we obtain (correct to $O(h^2)$ accuracy; as accurate as the compartment-based simulation)
 \begin{equation}\label{alg2gov} \D{p}{t} = D\frac{\partial^2 p(x,t)}{\partial x^2} + \lim_{\delta t\rightarrow 0}\frac{\mathbb{I}_{C_{-1}}(x)}{h\delta t} \int_t^{t+\delta t}\sum_j \delta(t'-t_j) - \sum_J\delta(t'-t_J)\mathrm{d}t'.
\end{equation}
In order to find equation (\ref{alg2gov}) it should be noted that
 \begin{equation}
 \mathbb{I}_{C_{-1}}(x)p(x,t) = \mathbb{I}_{C_{-1}}(x)\left(p(x_0,t)+p_x(x_0,t)(x-x_0)+\ldots\right)=\frac{\mathbb{I}_{C_{-1}}(x)}{h}\left(A_{-1}+O(h^2)\right),
 \end{equation}
 where $x_0$ is at the centre of $C_{-1}$ and $p_x(x_0,t) = O(h)$ (due to the no flux boundary condition placed on the PDE at the interface between the two subdomains). 
Evaluating the limit $\delta t \rightarrow 0$ we obtain
 \begin{equation}\label{alg2gov2}  \frac{\partial p(x,t)}{\partial t} = D\frac{\partial^2 p(x,t)}{\partial x^2} + \frac{\mathbb{I}_{C_{-1}}(x)}{h} \left(\sum_j \delta(t-t_j) - \sum_J\delta(t-t_J)\right).
\end{equation} 
If the number of particles within the pseudo-compartment, $A_{-1}$, becomes large then it will have a small relative variance and hence can effectively  be assumed to be deterministic (albeit not necessarily integer valued). 

Equation (\ref{alg2gov2}) is the motivating representation for Algorithm 2, which is valid for large $N_P$. Transfer events from $\Omega_P$ and $\Omega_C$ are calculated in the same way as Algorithm 1, however, their effect on the PDE domain is different. The source and sink terms in Algorithm 1, required for conditional probability changes in the PDE distribution as a result of periods of `no jump', are no longer part of Algorithm 2. In Algorithm 2, as particles transfer from $C_{-1}$ to $\Omega_C$ particles are withdrawn from the PDE assuming that they were from the pseudo-compartment in the same way that they are added when particles transfer from from $\Omega_C$ to $C_{-1}$ (as indicated in equation (\ref{alg2gov2})).
 
Unsurprisingly, for large $N_P$ we obtain a near deterministic hybrid simulation in both subdomains. Using equations (\ref{simp1}) and (\ref{simp2}), equation (\ref{alg2gov}) becomes
\begin{equation}\label{finalpde}
 \frac{\partial p(x,t)}{\partial t} = D\frac{\partial^2 p(x,t)}{\partial x^2} + \frac{\mathbb{I}_{C_{-1}}(x)D}{h^2}\left( A_1 - A_{-1} \right).
\end{equation}
The last two terms in equation (\ref{finalpde}) represents a lattice approximation to a free diffusive flux over the interface, a desired outcome of Algorithm 2.

\begin{figure}[h!!!!!!!!!!!!!!!!!!!!]
\begin{center} 
\includegraphics[width=\columnwidth]{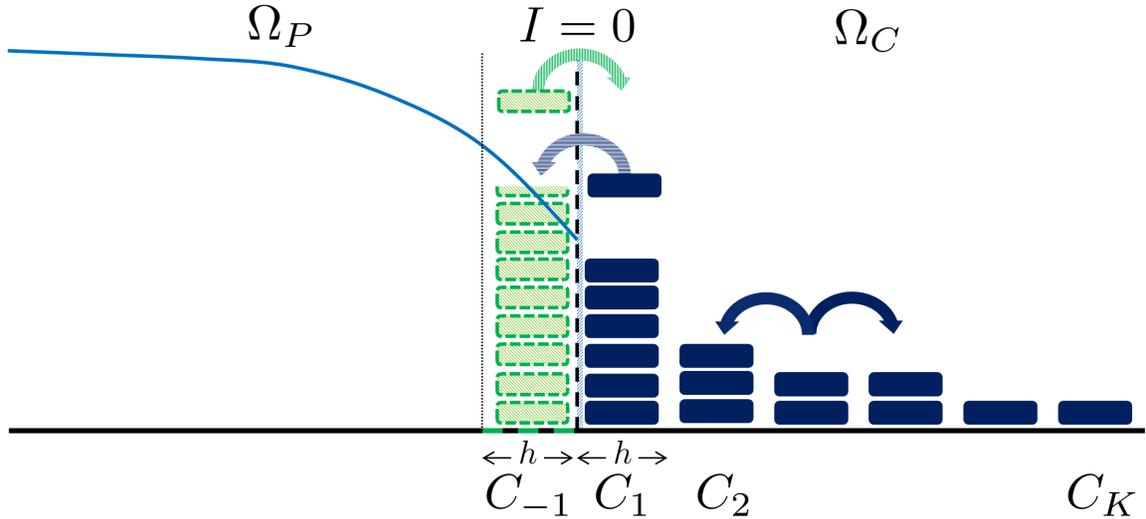}
\end{center}
\caption{A cartoon schematic illustrating the most important features of Algorithm 2 in one dimension. In $\Omega_C$, to the right of the interface, particles are represented as individuals in compartments (dark blue rectangles) with each individual adding $1/h$ to the density in that compartment. Particles can jump to neighbouring compartments within $\Omega_C$ (dark blue block arrows). In $\Omega_P$, to the left of the interface, density is represented by a `continuous' function (light blue curve).  Particles in the left-most compartment in $\Omega_C$ (adjacent to the interface) can jump into the PDE regime (blue horizontally striped arrow) where they are added as an indicator function to the PDE solution in the region $[-h,0)$ - the pseudo-compartment (green dashed line). Pseudo-particles (hatched green rectangles with dashed borders) corresponding to the mass of the PDE in the pseudo-compartment are allowed to jump right (green vertically striped arrow) into the first compartment in $\Omega_C$. When a 
particle leaves the pseudo-compartment an indicator function of mass corresponding to one particle is removed from the PDE in that region.}
\label{figure:cartoon_schematic}
\end{figure}

 A cartoon schematic illustrating the simplified algorithm for large copy numbers (Algorithm 2) in one dimension is given in Figure \ref{figure:cartoon_schematic}.
 
\begin{table}[h!!!!!!!!!!!!!]
\begin{tabular}{p{\textwidth}}
\hline
\begin{enumerate}[(i)]
 \item Initialise the time, $t=t_0$ and set the final time, $t_{final}$. Specify the PDE-update time-step $dt_{P}$ and initialise the next PDE time-step to be $t_P=t+dt_{P}$.
 \item Initialise the number of particles in each compartment in $\Omega_C$, $A_i$ for $i=1,\dots, K$, and the distribution of probability density in $\Omega_P$, $p(x,t_0)$, for $x\in\Omega_P$.
 \item \label{item:calculate_the_propensity_functions} Calculate the propensity functions for diffusion between the compartment regimes as $\alpha_{i,j}=A_iD/h^2$ for $i=1\dots K$ and $j=M+1, M+2$ and for reactions as $\alpha_{i,j}$ for $i=1\dots K$ and $j=1,\dots, M$ using the usual mass action kinetics.
 \item Calculate the propensity function for diffusion from the pseudo-compartment, $C_{-1}$, in $\Omega_P$ into the adjacent compartment in $\Omega_C$: $\alpha^*=D\int_{C_{-1}}p(x,t)\ud x/h^2$ (equation \eqref{ghostjump}).
 \item Calculate the sum of the propensity functions, $\alpha_0=\sum^K_{i=1}\sum^{M+2}_{j=1}\alpha_{i,j}+\alpha^*$.
 \item Determine the time for the next `compartment-based' event $t_C=t+\tau$, where $\tau$ is given by equation \eqref{equation:tau_choice}.
 \item \label{item:update_particle_numbers_in_compartments} If $t_C<t_P$ then the next compartment-based event occurs:
 \begin{enumerate}[(a)]
 \item determine which event occurs according to the method described in the text (see \citet{gillespie1977ess}).
 \item If the event corresponds to $\alpha_{i,j}$ for $i=1\dots K$ and $j=M+1,M+2$ then move a particle from compartment $i$ in the direction specified by $j$. If the particle crosses the interface into pseudo-compartment $C_{-1}$ then add a particle's worth of mass to the region $C_{-1}$ i.e. $p(x,t+\tau)=p(x,t)+\mathbb{I}_{C_{-1}}/h$. Here $\mathbb{I}_{C_{-1}}$ is the indicator function which takes the value 1 in $C_{-1}$ and 0 elsewhere. 
  \item \label{item:remove_mass_from_PDE} If the event corresponds to propensity function $\alpha^*$ and $p(x,t)>1/h \quad \text{ for all }\quad x\in C_{-1}$ then place a particle in $C_1$. Remove a particle's worth of mass from the PDE solution in the  region $C_{-1}$ i.e. $p(x,t+\tau)=p(x,t)-\mathbb{I}_{C_{-1}}/h$.
  \item update the current time, $t=t_C$. 
 \end{enumerate}
\item If $t_P<t_C$ the the PDE regime is updated:
\begin{enumerate}[(a)]
\item Update the PDE solution according to the numerical method described in Section \ref{section:PDE_based_modelling} using $p(x,t)$ as the previous value of the solution.
 \item Update the current time, $t=t_P$ and set the time for the next PDE update step to be $t_P=t_P+dt_P$.
\end{enumerate}
\item If $t\leq t_{final}$, return to step \eqref{item:calculate_the_propensity_functions}.

Else end.
\end{enumerate}\\
\hline
\end{tabular}
\caption{Algorithm for the coupling of the compartment-based regime with the PDE-based regime in the case of high copy numbers.}
\label{table:algorithm_2}
\end{table}

Table \ref{table:algorithm_2} describes the algorithm used to interface the two regimes. Note that when removing a pseudo-particle from the PDE regime in step \eqref{item:update_particle_numbers_in_compartments} (c)
we enforce the condition that $p(x,t)>1/h \quad \text{ for all }\quad x\in C_{-1}$ to ensure that when we remove an indicator function corresponding to a particle's worth of mass that we do not make the PDE solution negative. In practise this condition will rarely be enforced since the PDE region should be used to describe regions of the domain that have sufficiently high density that the individual-based model becomes inefficient. We note, however, that with a fixed interface this will not always be possible and that this restriction may lead to slight inaccuracies in the solution.

We have shown that, theoretically, Algorithm 1 simplifies to Algorithm 2 in the case of large $N_P$. In the following section, we will demonstrate this empirically on a number of simple test problems. We will also demonstrate the strong agreement between the hybrid simulations of both algorithms and known exact analytic solutions.

\section{Results}
In this section we assesses the accuracy of our proposed algorithms in simulating three straightforward test problems of increasing complexity. In all the example simulations which follow we choose $dt_p$, $\Delta x=0.005$ and $h=0.05$, so that for every compartment there are 10 PDE mesh points. These parameters will be systematically varied later in this section in order to determine the error behaviour of the algorithm.

\subsection{Test Problem 1: Uniform particle distribution}

In Test Problem 1 we demonstrate that our proposed algorithms are capable of maintaining a steady state distribution of particles. We consider the diffusion of non-interacting particles on a domain with reflective boundaries which will maintain a steady state distribution. We assume that all $N$ particles are initially uniformly distributed across the domain and move with diffusion coefficient $D=1/400$, in non-dimensionalised units. The PDE that we solve in $\Omega_P$ is the diffusion equation with zero flux boundary conditions at either end.

\begin{figure}[h!!!!!!!]
\begin{center}
\subfigure[]{
\includegraphics[width=0.31\columnwidth]{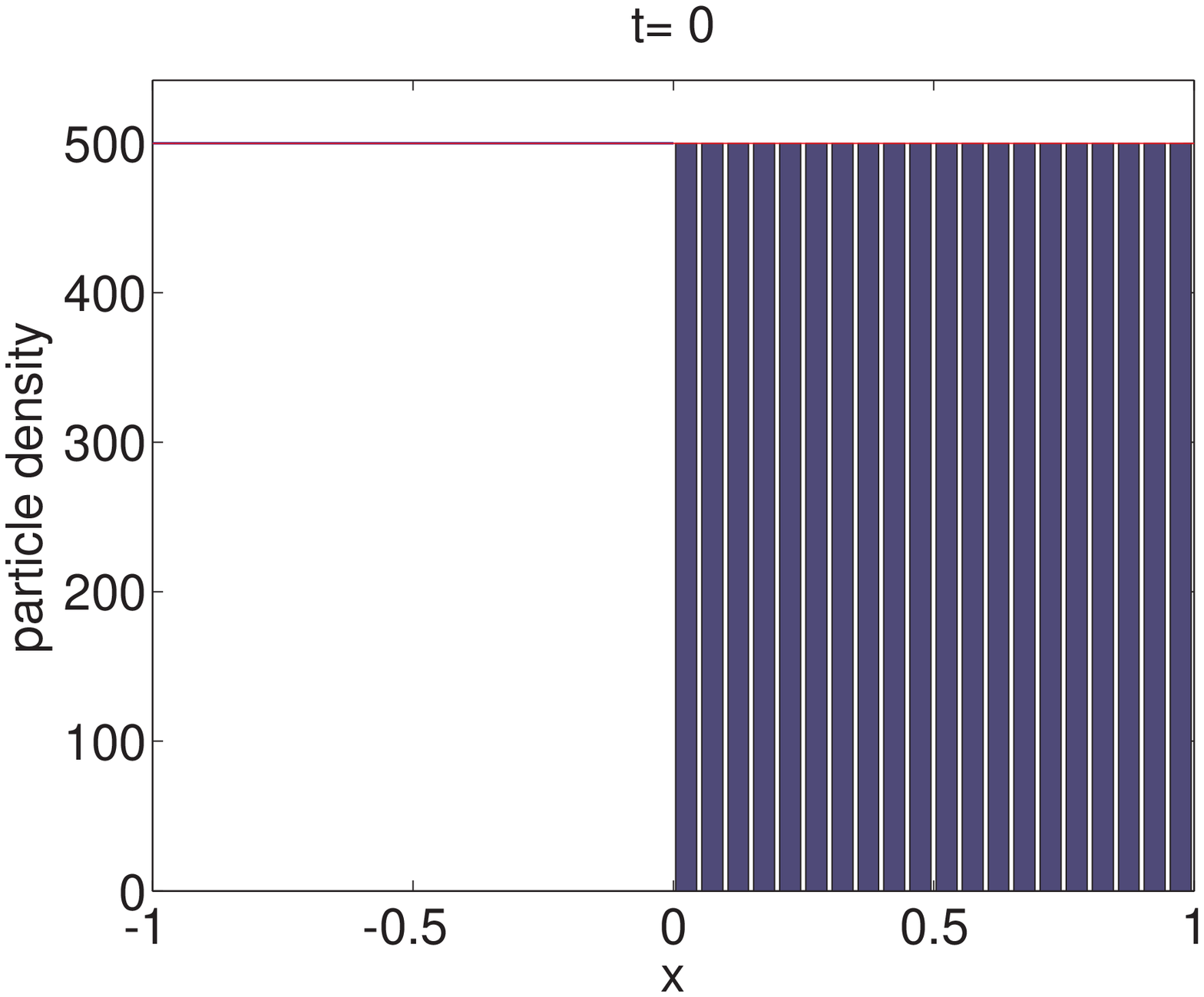}
\label{figure:uniform_initial_condition}
}
\subfigure[]{
\includegraphics[width=0.31\columnwidth]{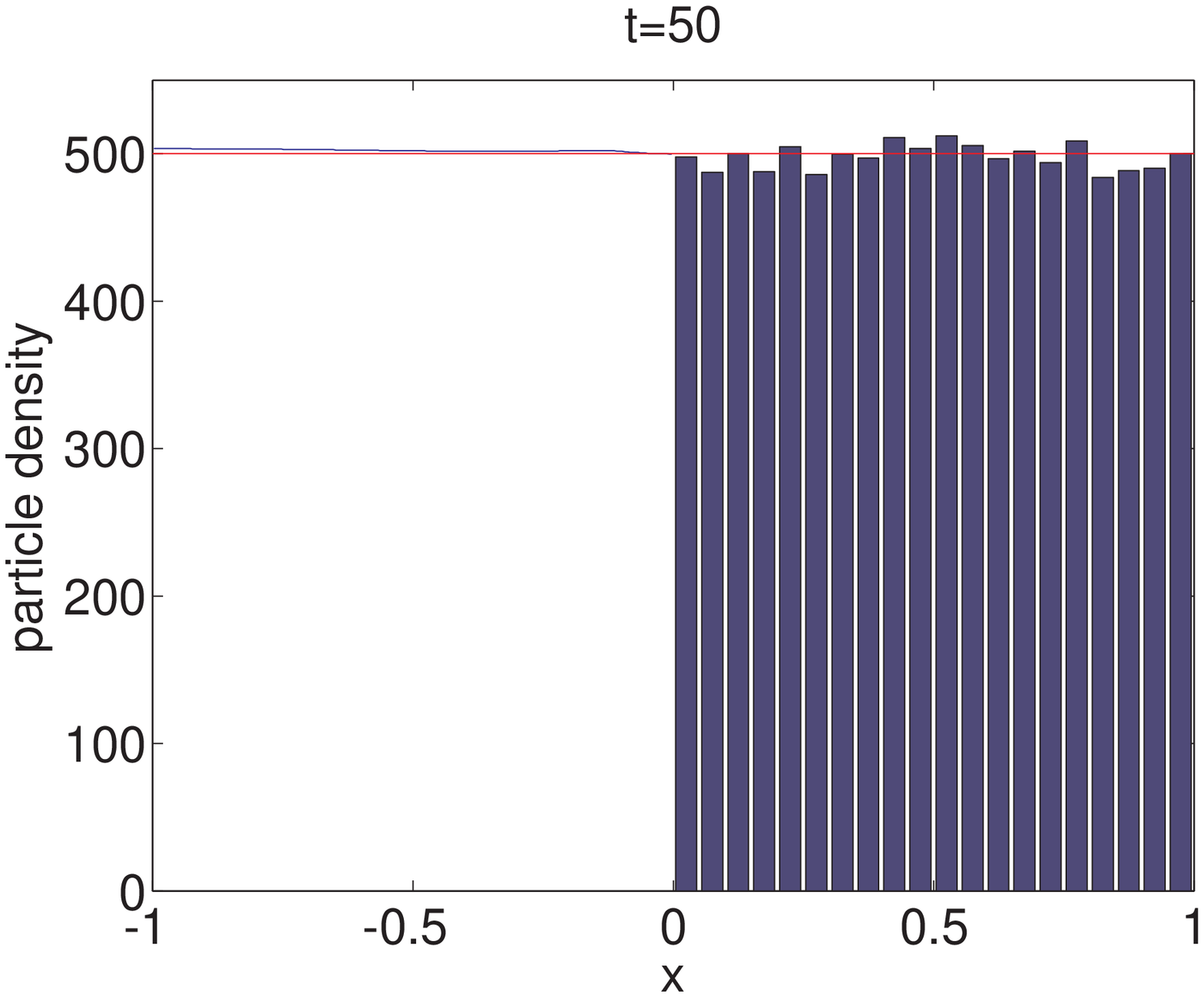}
\label{figure:uniform_algorithm_1}
}
\subfigure[]{
\includegraphics[width=0.31\columnwidth]{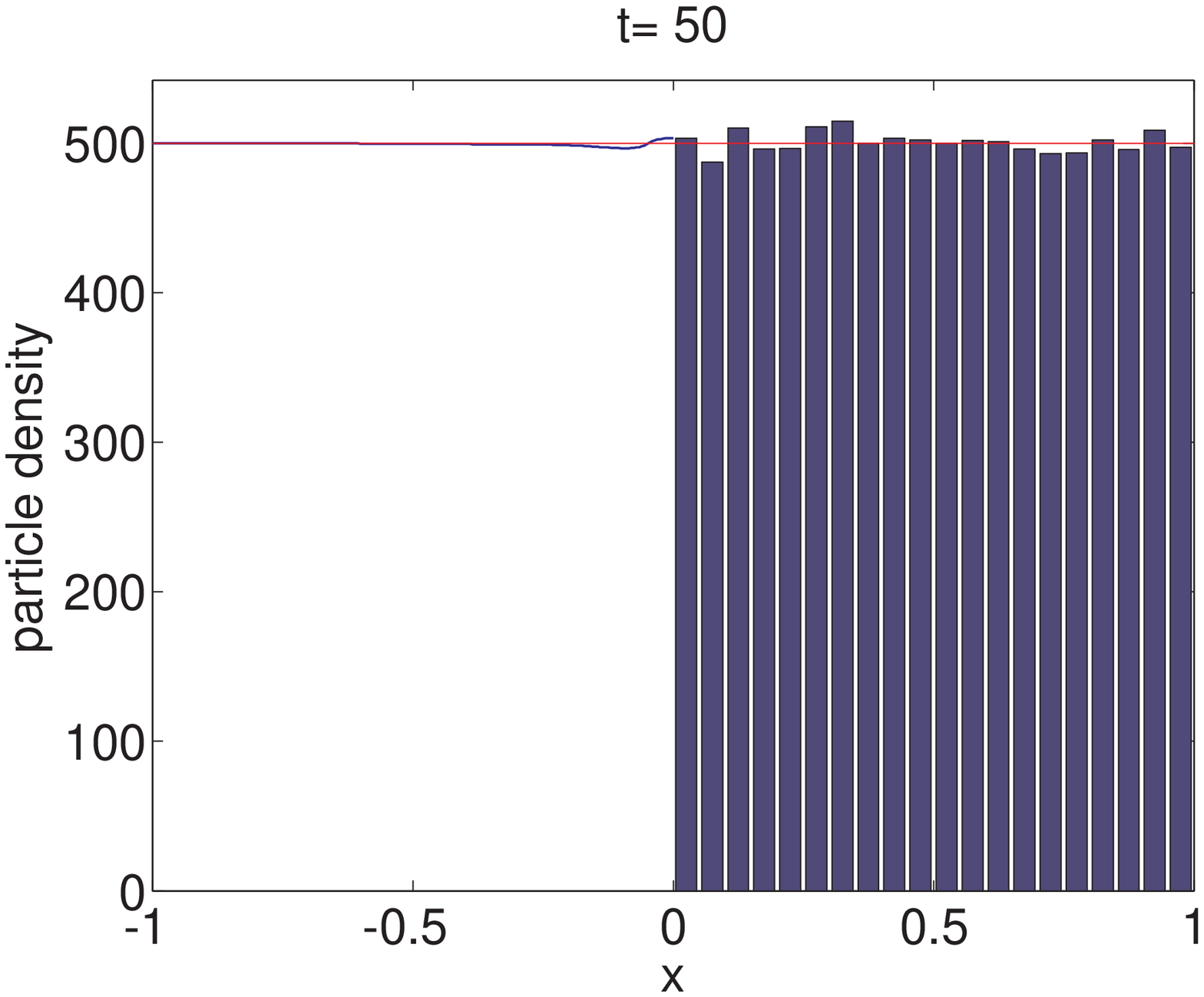}
\label{figure:uniform_algoritm_2}
}
\end{center}
\caption{Maintenance of  \subref{figure:uniform_initial_condition} an initially uniform steady state by \subref{figure:uniform_algorithm_1} Algorithm 1  and \subref{figure:uniform_algoritm_2} Algorithm 2. In each subfigure the blue curve is the solution of the PDE in $\Omega_P$, the blue histograms represent the concentration of particles in each of the compartments in $\Omega_C$ and the red curve represents the analytic solution of the mean-field model. Note that the blue curve is difficult to see since, for the most-part, it lies underneath the analytic solution. The number of particles is $N=1000$. The simulation results are averaged over 100 repeats.}
\label{figure:uniform}
\end{figure}

Figure \ref{figure:uniform} demonstrates that a uniform steady state is maintained by both algorithms. The disparity between the PDE solution and the analytic solution close to the interface in Figure \ref{figure:uniform} \subref{figure:uniform_algoritm_2} is transient and oscillates either side of the analytic solution as time progresses (see Supplementary Material (SM) Movies M1 and M2 corresponding to Algorithms 1 and 2, respectively). 

\subsection{Test Problem 2: Particle redistribution}

As a further example of the propriety of our algorithms we consider a situation in which particles are initially uniformly distributed in either $\Omega_P$ or in $\Omega_C$ with no particles in the complementary region. We note that the situation in which all the particles are in $\Omega_C$ and hence the concentration in the compartment-based region is much higher than that in the PDE region is not the ideal situation in which to employ our algorithm. However, it is important to show that our algorithm is robust to such situations. In physical terms each of these simulations corresponds to particles, artificially kept in one half of a box, being allowed to diffuse into the other half of the box.

The mean-field model corresponding to this situation can be formulated as an initial-boundary-value problem:
\begin{equation}
 \D{\rho}{t}=D\DD{\rho}{x}, \text{ for }   x\in [-1,1],
\end{equation}
with zero-flux boundary conditions
\begin{equation}
 \D{\rho}{x}\Big|_{x=-1,1}=0,
\end{equation}
and initial conditions
\begin{equation}
\rho(x,0)=H(-x) \text{ for }  x\in [-1,1],\label{equation:PDE_initial_condition}
\end{equation}
or 
\begin{equation}
\rho(x,0)=H(x) \text{ for }  x\in [-1,1],\label{equation:compartment_initial_condition}
\end{equation}
depending on which side of the box we wish to initialise the particles. 
This equation can be solved using separation of variables and Fourier series.
For initial condition \eqref{equation:PDE_initial_condition} the solution is given by:
\begin{equation} \rho(x,t)=\frac{1}{2}-\sum_{n=1}^{\infty}\frac{2}{(2n-1)\pi}(-1)^n\cos\left((2n-1)\pi\left(\frac{x+1}{2}\right)\right)\exp\left(-\frac{(2n-1)^2\pi^2Dt}{4}\right).\label{equation:heaviside_analytical_solution}
\end{equation}
For initial condition \eqref{equation:compartment_initial_condition} the minus in front of the sum is replaced with a plus.

\begin{figure}[h]
\begin{center}
\subfigure[]{
\includegraphics[width=0.31\columnwidth]{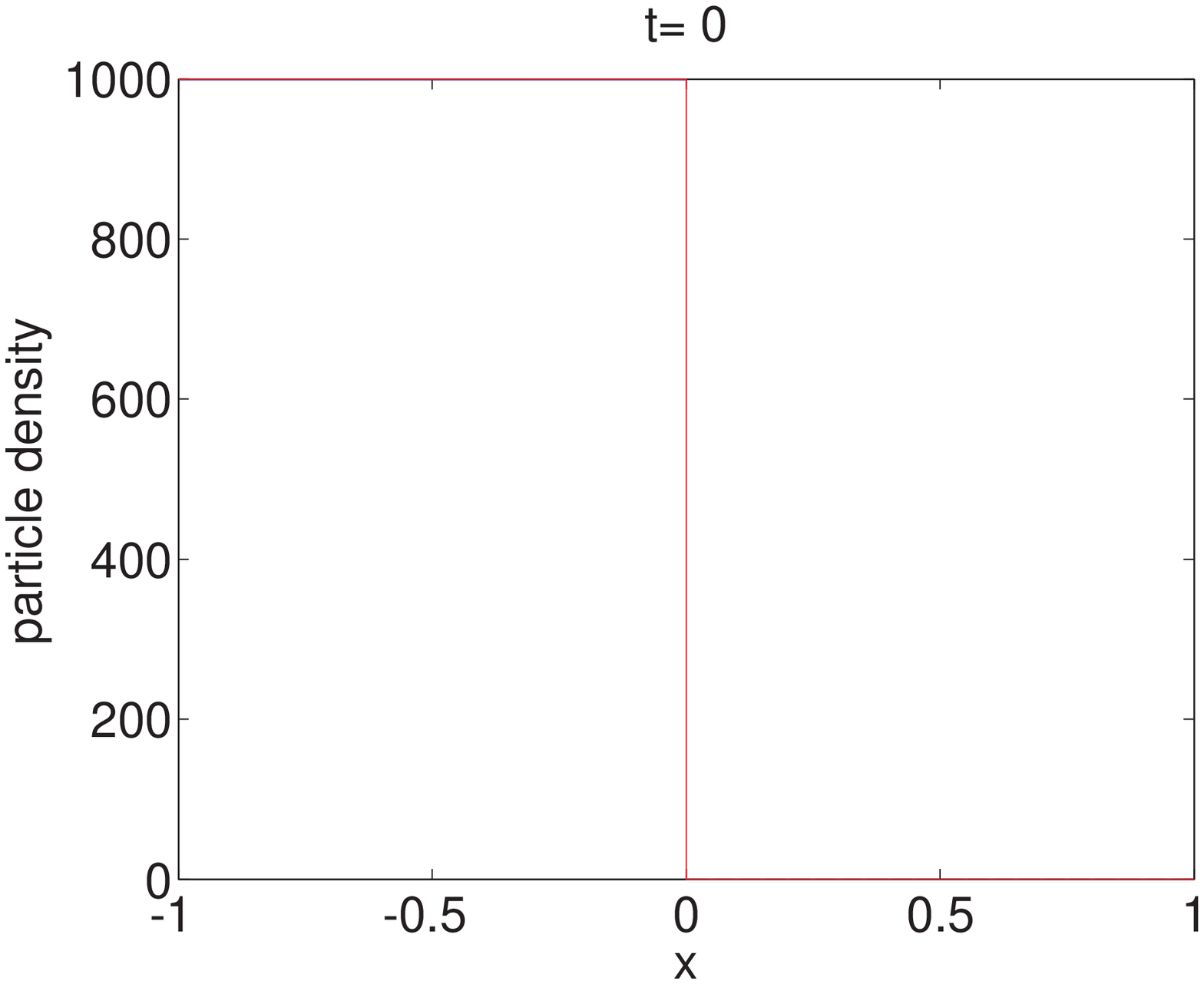}
\label{figure:heaviside_initial_condition}
}
\subfigure[]{
\includegraphics[width=0.31\columnwidth]{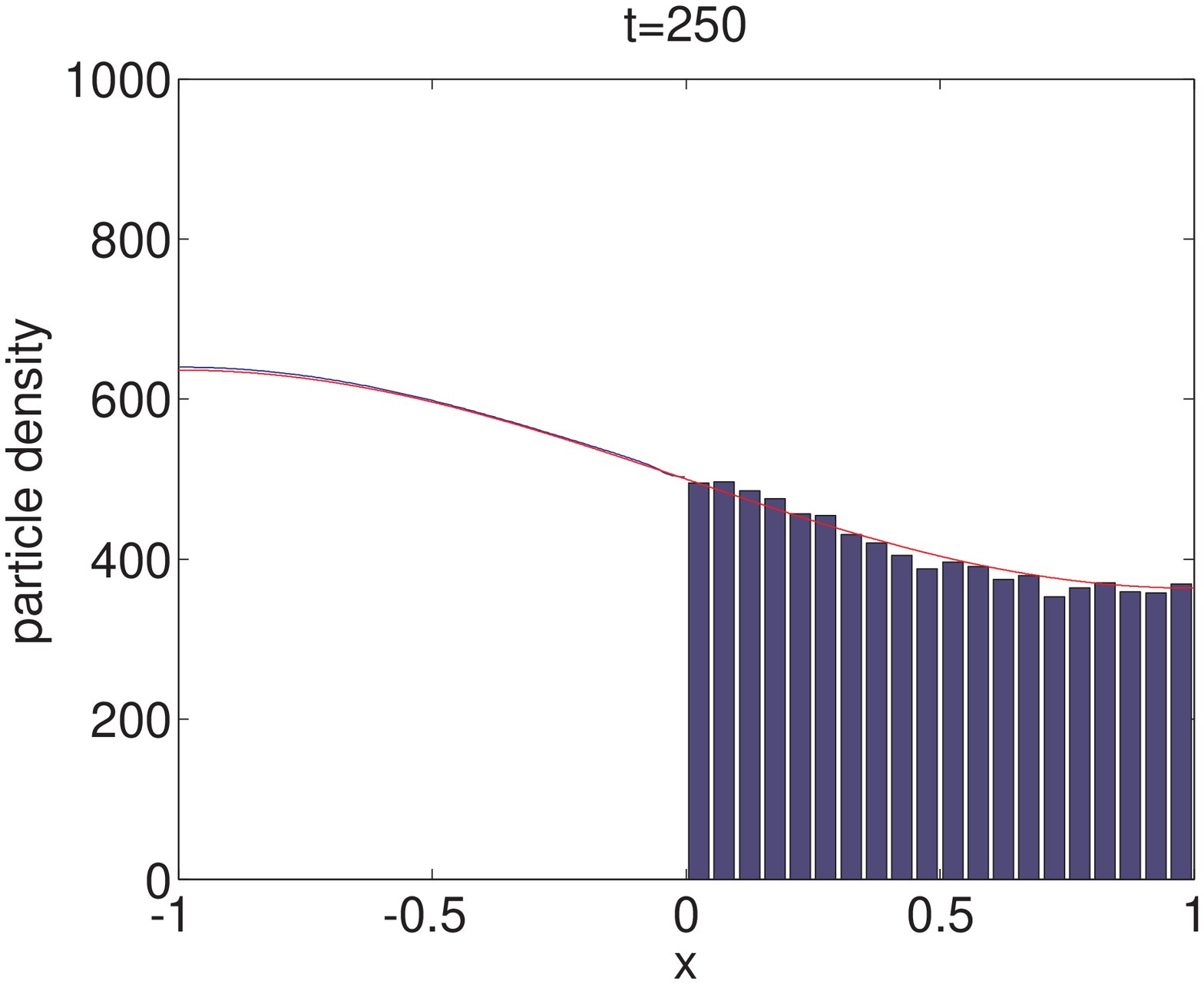}
\label{figure:heaviside_250_algorithm_1}
}
\subfigure[]{
\includegraphics[width=0.31\columnwidth]{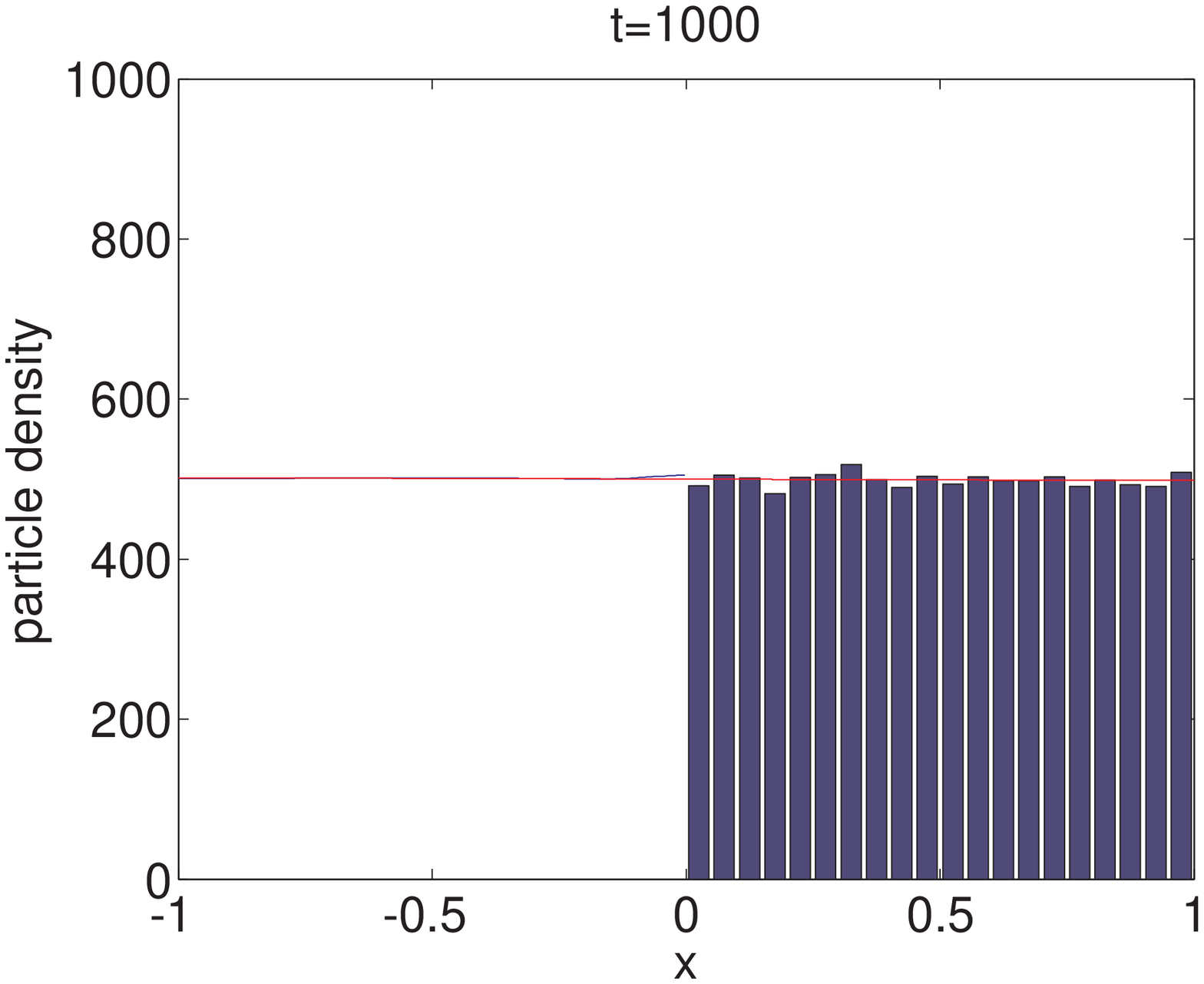}
\label{figure:heaviside_1000_algorithm_1}
}
\subfigure[]{
\includegraphics[width=0.31\columnwidth]{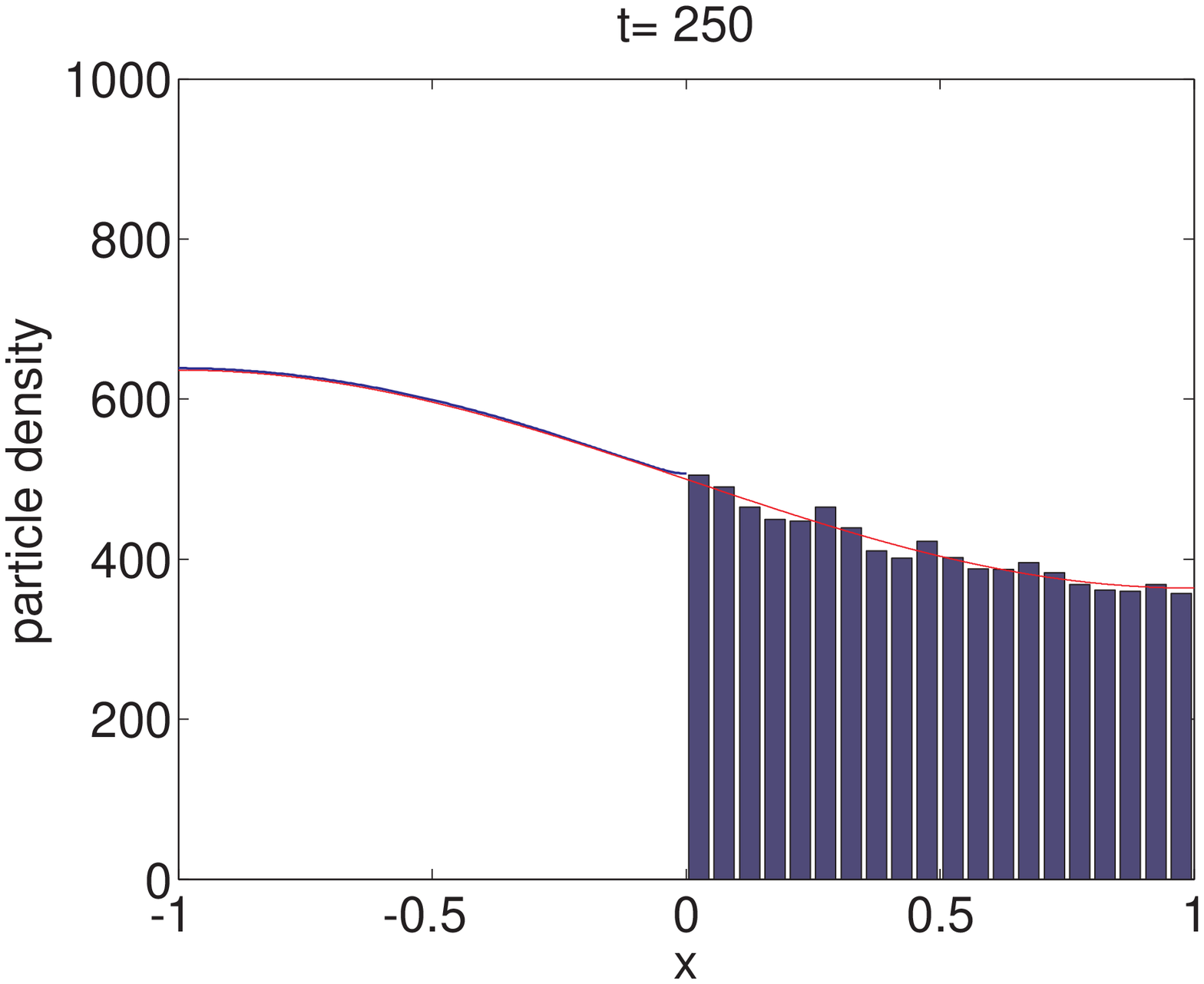}
\label{figure:heaviside_250_algorithm_2}
}
\subfigure[]{
\includegraphics[width=0.31\columnwidth]{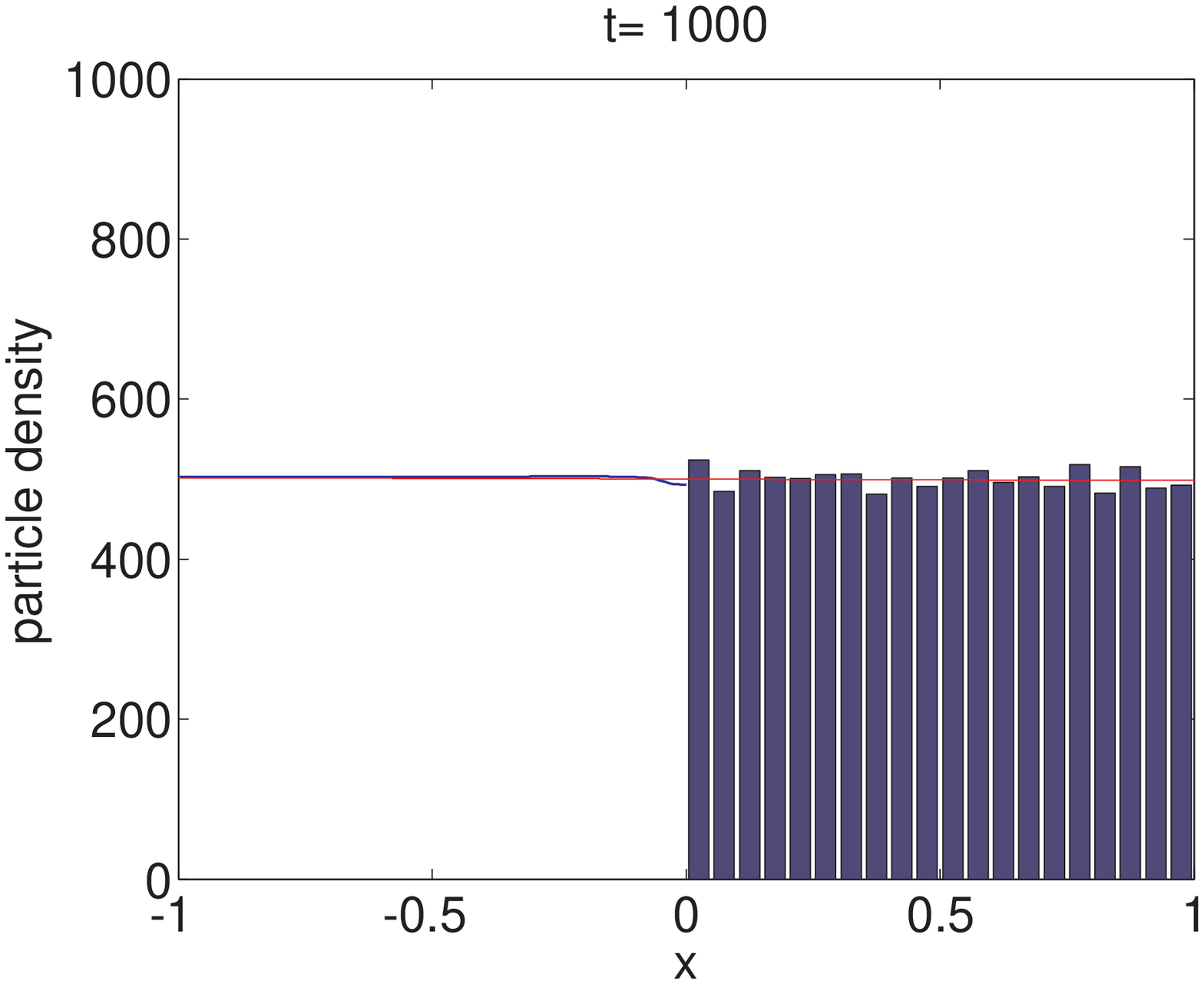}
\label{figure:heaviside_1000_algorithm_2}
}
\end{center}
\caption{Particles initially in the left-hand half of the domain \subref{figure:heaviside_initial_condition} spread diffusively (with diffusion coefficient $D=1/400$) throughout the domain eventually reaching steady state. By $t=1000$ the system is assumed to have reached steady state. Panels \subref{figure:heaviside_250_algorithm_1} and \subref{figure:heaviside_1000_algorithm_1} show the particle density profile according to Algorithm 1 at times 250 and 1000 respectively. Panels \subref{figure:heaviside_250_algorithm_2} and \subref{figure:heaviside_1000_algorithm_2} show the same comparison for Algorithm 2. Figure descriptions and particle numbers are as in Fig \ref{figure:uniform}. The simulation results are averaged over 100 repeats.}
\label{figure:heaviside_PDE_IC}
\end{figure}

In Figure \ref{figure:heaviside_PDE_IC} we compare the analytically derived density to the density produced using our algorithms given initial condition \eqref{equation:PDE_initial_condition}. We see a favourable comparison between the solution according to our algorithms and the mean-field solution. A comparison of the densities through time for Algorithm 1 and 2 can be seen in Movies M3 and M4 of the SM, respectively. In Movies M5 and M6 of the SM we present the same comparison but for a single realisation of Algorithms 1 and 2 respectively. In Figure 1
of the SM we compare the particle density of the two algorithms against the analytic solution of the mean-field model for initial condition \eqref{equation:compartment_initial_condition}. The comparison is again favourable. 

\subsection{Test Problem 3: A morphogen gradient formation model}

In this final test problem we model the formation of a morphogen gradient by considering basic first order reactions.  Morphogen is produced at $x=-1$ with rate $D\lambda$. We implement this production as a flux boundary condition in the PDE region of the model:
\begin{equation}
\D{p}{x}\Big|_{x=-1}=-\lambda.\label{equation:morphogen_flux_boundary_condition}
\end{equation}
We also allow morphogen to decay uniformly across the domain with rate $\mu$ and we implement a zero-flux boundary condition at $x=1$. These zeroth order (production) and first order (degradation) reactions allow for the description of the mean dynamics of the fully stochastic model by a PDE:
\begin{equation}
 \D{\rho}{t}=D\DD{\rho}{x}-\mu \rho.\label{equation:morphogen_gradient_equation}
\end{equation}
This is the mean-field PDE which we expect the density to satisfy across the whole domain. The analytic solution for the expected evolution of the density for this system (with boundary condition analogous to equation \eqref{equation:morphogen_flux_boundary_condition}) can therefore be found explicitly as
\begin{equation}
\rho(x,t)=\rho_{st}(x)+\rho_{u}(x,t),\label{equation:mean_field_solution}
\end{equation}
where
\begin{equation} \rho_{st}(x)=\lambda\sqrt{\frac{D}{\mu}}\frac{\cosh\left(\sqrt{\mu/D}(1-x)\right)}{\cosh\left(2\sqrt{\mu/D}\right)},
\end{equation}
gives the steady state of the morphogen gradient and
\begin{equation}
 \rho_{u}(x,t)=-\frac{\lambda D}{2\mu}\tanh\left(2\sqrt{\frac{\mu}{D}}\right)\left\{\exp(-\mu t)+2\displaystyle\sum_{n=1}^{\infty}\frac{\cos(n\pi(x+1)/2)}{1+D(n\pi)^2/4\mu}\exp\left(-\left(\frac{D (n\pi)^2}{4}+\mu\right)t\right)\right\}.\label{equation:unsteady_solution}
\end{equation}
Since equation \eqref{equation:morphogen_gradient_equation} is the PDE we would expect the mean-field density of the  fully stochastic model to satisfy, we use this as the PDE we solve in $\Omega_P$ in our hybrid model. In $\Omega_C$ we implement the usual diffusive `reactions' and additional mono-molecular decay reactions in each compartment and a zero flux boundary condition at $x=1$.

\begin{figure}[h!!!!!!!!!!!!!!!!!]
\begin{center}
\subfigure[]{
\includegraphics[width=0.31\columnwidth]{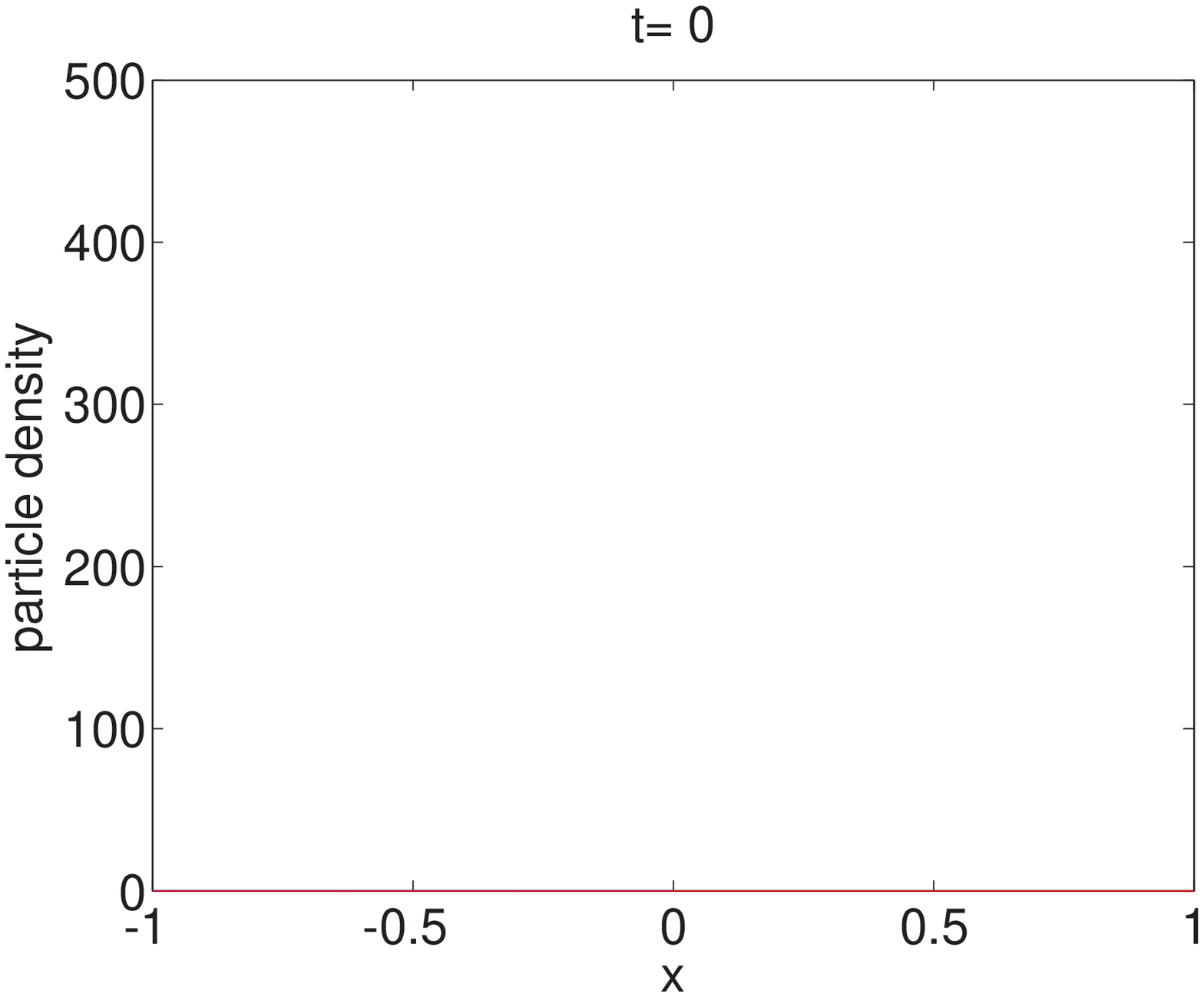}
\label{figure:morphogen_initial_condition}	
}
\subfigure[]{
\includegraphics[width=0.31\columnwidth]{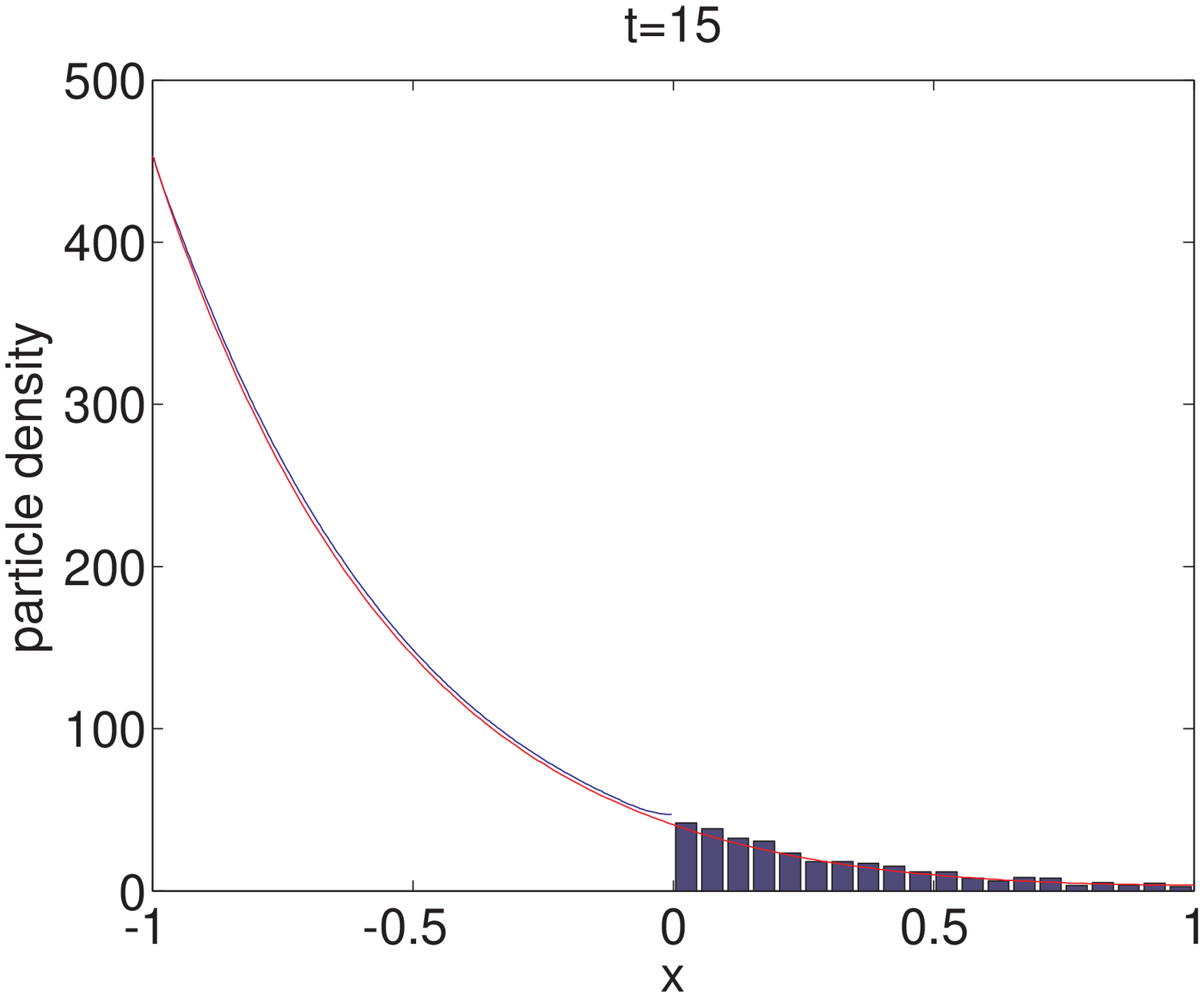}
\label{figure:morphogen_15_algorithm_1}
}
\subfigure[]{
\includegraphics[width=0.31\columnwidth]{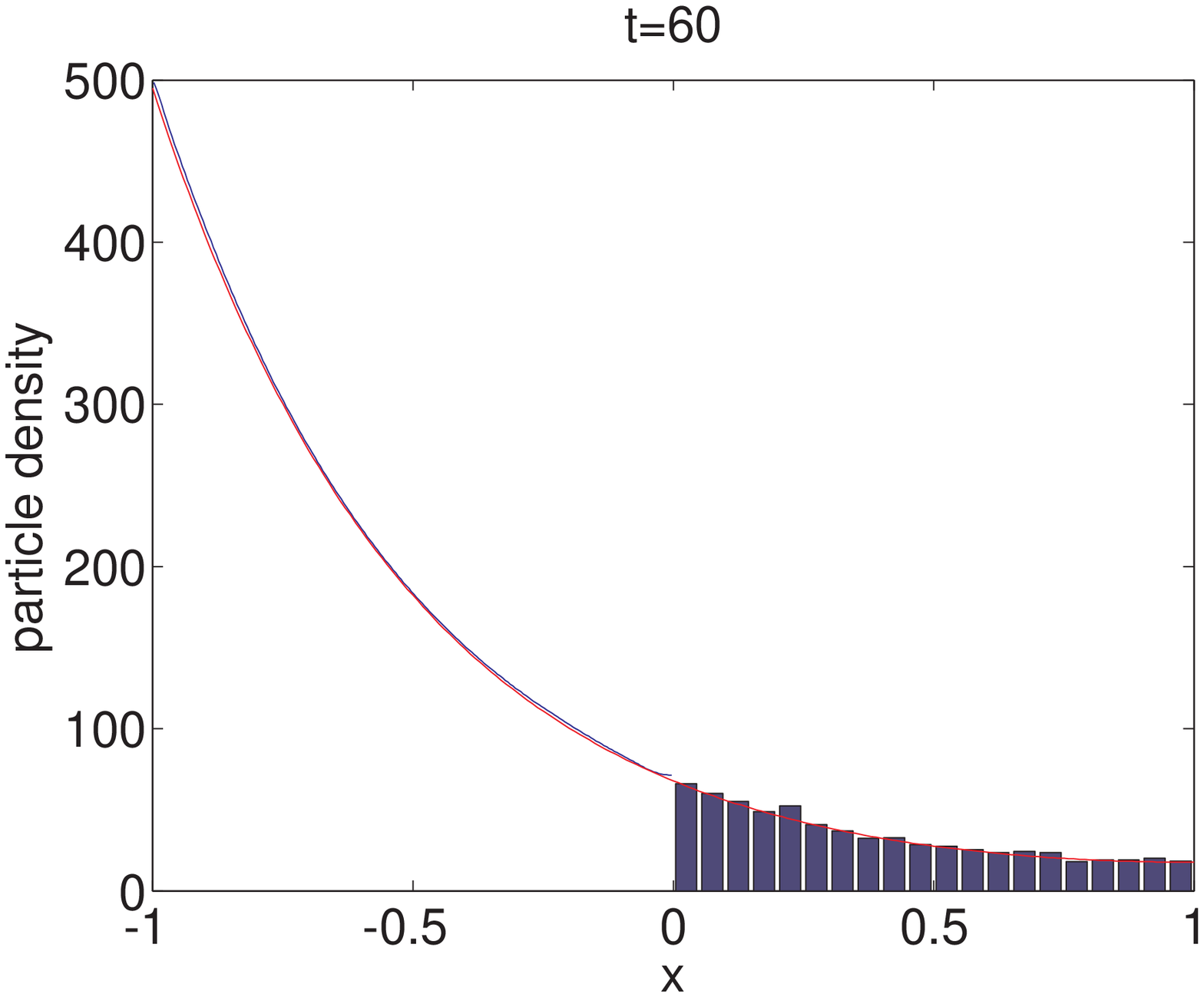}
\label{figure:morphogen_60_algorithm_1}
}
\subfigure[]{
\includegraphics[width=0.31\columnwidth]{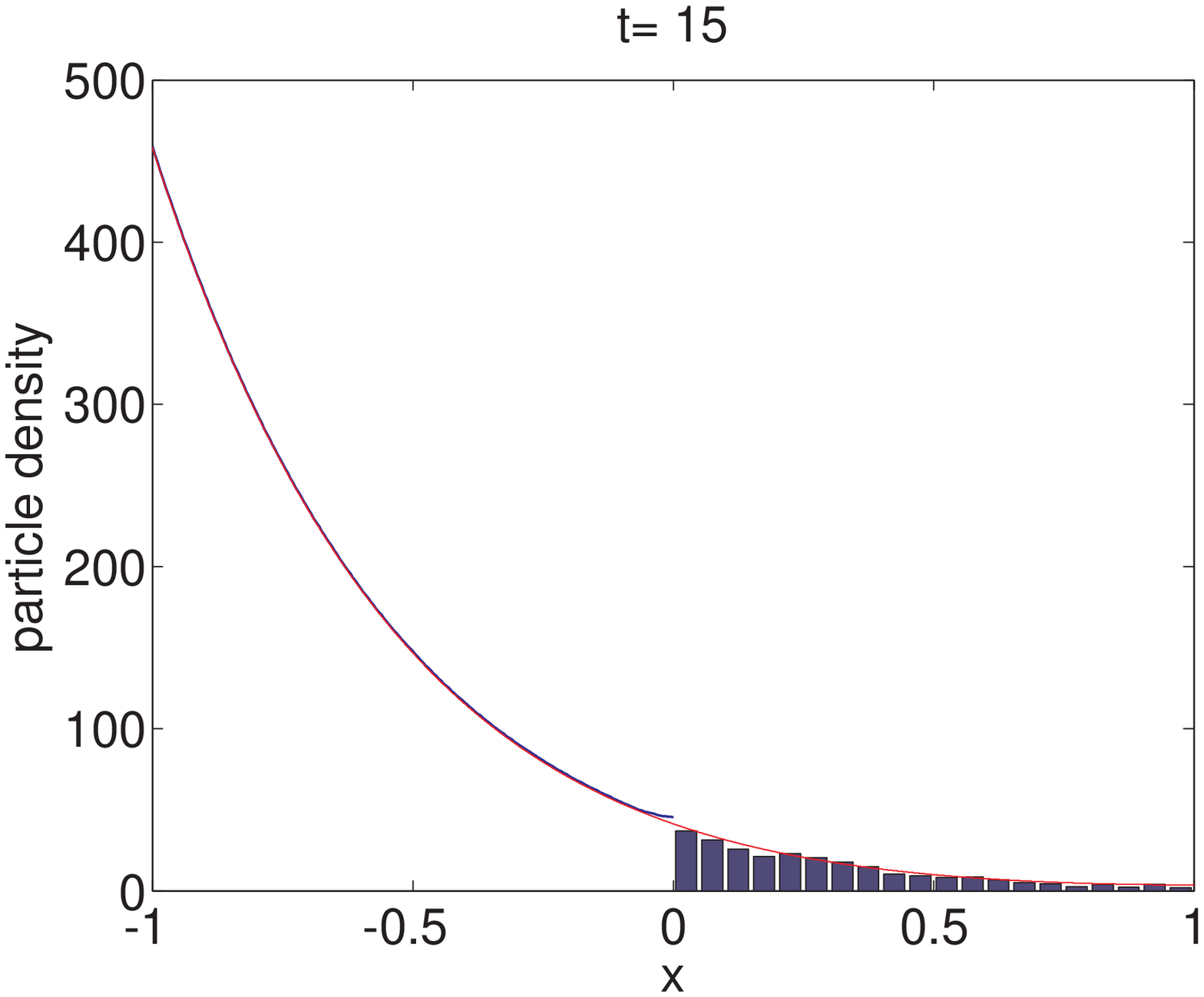}
\label{figure:morphogen_15_algorithm_2}
}
\subfigure[]{
\includegraphics[width=0.31\columnwidth]{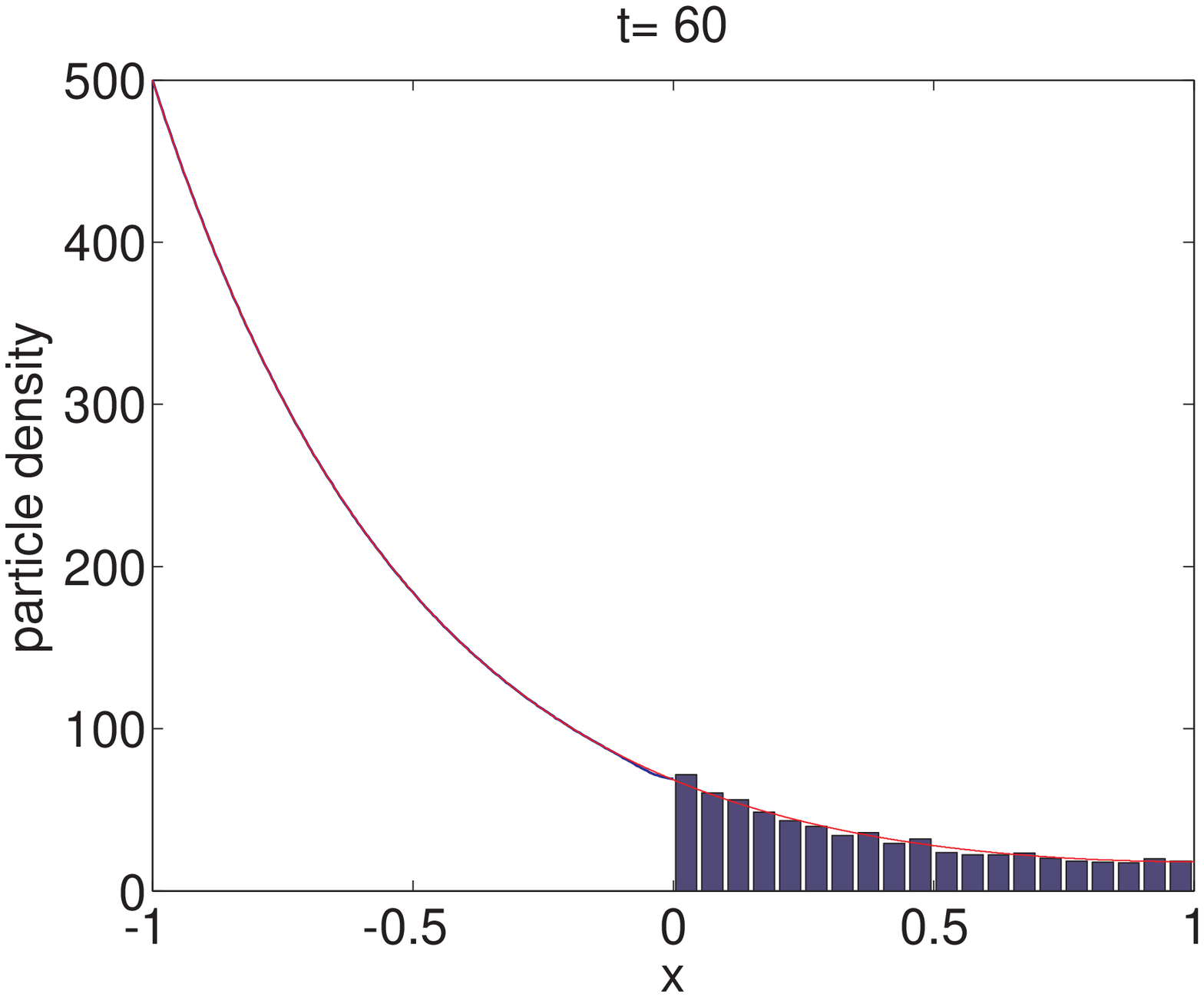}
\label{figure:morphogen_60_algorithm_2}
}
\end{center}
\caption{A morphogen gradient forms though a combination of influx at the left-hand boundary and decay of morphogen particles throughout the domain. Panel \subref{figure:morphogen_initial_condition} shows the initially empty domain. By $t=60$ the system is assumed to have reached an approximately steady state. Panels \subref{figure:morphogen_15_algorithm_1} and \subref{figure:morphogen_60_algorithm_1} show the particle density profile according to Algorithm 1 at times 15 and 60 respectively. Panels \subref{figure:morphogen_15_algorithm_2} and \subref{figure:morphogen_60_algorithm_2} show the same comparison for Algorithm 2. Figure descriptions are as in Fig \ref{figure:uniform}. The simulation results are averaged over 100 repeats.}
\label{figure:morphogen_gradient}
\end{figure}

In Figure \ref{figure:morphogen_gradient} we compare the analytically derived density to that produced using our algorithms given an initially empty domain. We increase the value of the diffusion coefficient by a factor of 10  (in comparison to previous simulations) to $D=10\times h^2=1/40$ (in order to achieve a sensible steady state profile). We choose $\lambda=1000$ and $\mu=0.1$, giving a rate of influx of particles into the domain of $D\lambda=25$. As in previous examples, we see a favourable comparison between the solution according to our hybrid algorithms and the expected mean-field solution given by equations \eqref{equation:mean_field_solution}-\eqref{equation:unsteady_solution}. A comparison of the densities through time for Algorithms 1 and 2 can be seen in Movies M7 and M8 of the SM, respectively.

A more quantitative study of the similarities between the solution of our algorithms and the analytic solutions is presented in the next section.

\section{Error analysis}
In this section we revisit the comparison between the mean particle densities predicted by our hybrid algorithm and the expected values given by the corresponding deterministic model. It is our aim in developing our algorithms that the error between the hybrid methods and the deterministic solution should be of the same order of magnitude as the accuracy associated with the individual numerical simulation techniques. In particular we developed our algorithms in order that the stochastic model should not `see' (i.e. be influenced by) the interface. Since it is well established that the descriptions of particle density in $\Omega_P$ and $\Omega_C$ are valid representations of diffusion, we address the question of `whether our algorithms can correctly simulate the flux over the interface?' In order to test this we compare \begin{equation}                                                                                                                                                                                 
                                                                                                                                                                                                                                                                                                                                                                                                                                                                                                                                                                                                                                                                                                                                                                                                                                                                                                                                                           N_{DC}(t)=\int_0^1\rho(x,t)\ud x,
 \end{equation}
the expected number of particles in $\Omega_C$ in the deterministic model to  
\begin{equation}
N_{HC}(t)=\displaystyle\sum_{i=1}^{K}\bar{A}_i,
\end{equation}
the expected number of particles in $\Omega_C$ in the hybrid models. Here $\bar{A}_i$ represents the mean number of particles in the $i^{th}$ compartment averaged over 100 repeats of the hybrid algorithms.
For completeness we also compare 
\begin{equation}
N_{DP}(t)=\int_{-1}^0\rho(x,t)\ud x,
 \end{equation}
the expected number of particles in $\Omega_P$ in the deterministic model to
\begin{equation}
N_{HP}(t)=\int_{-1}^0 p(x,t)\ud x,
\end{equation}
the expected number of particles in $\Omega_P$ in the hybrid models.
 
\begin{figure}[h!!!!!!!!!!!!!!!!!]
\begin{center}
\subfigure[]{
\includegraphics[width=0.45\columnwidth]{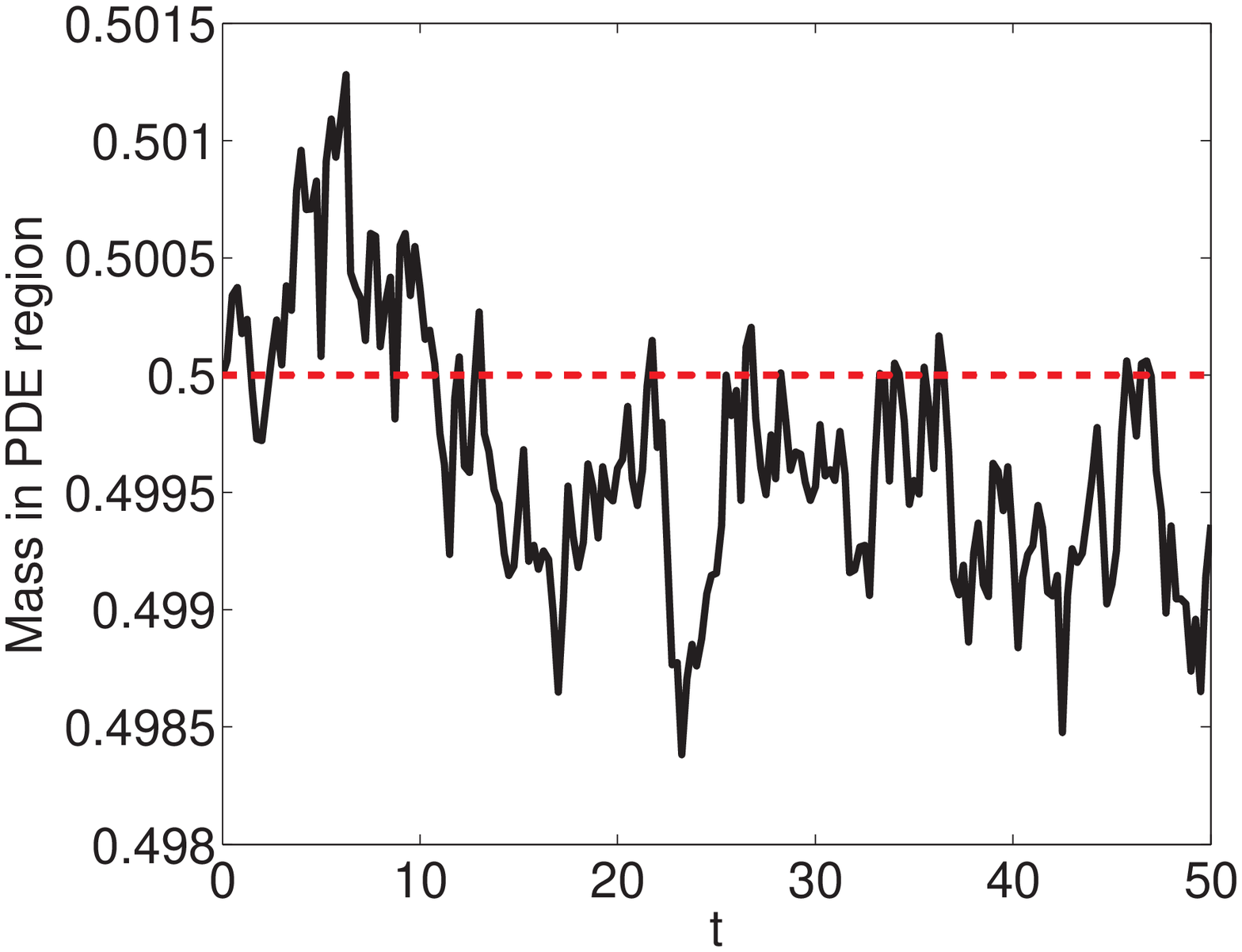}
\label{figure:uniform_PDE_mass_and_error_alg2}
}
\subfigure[]{
\includegraphics[width=0.45\columnwidth]{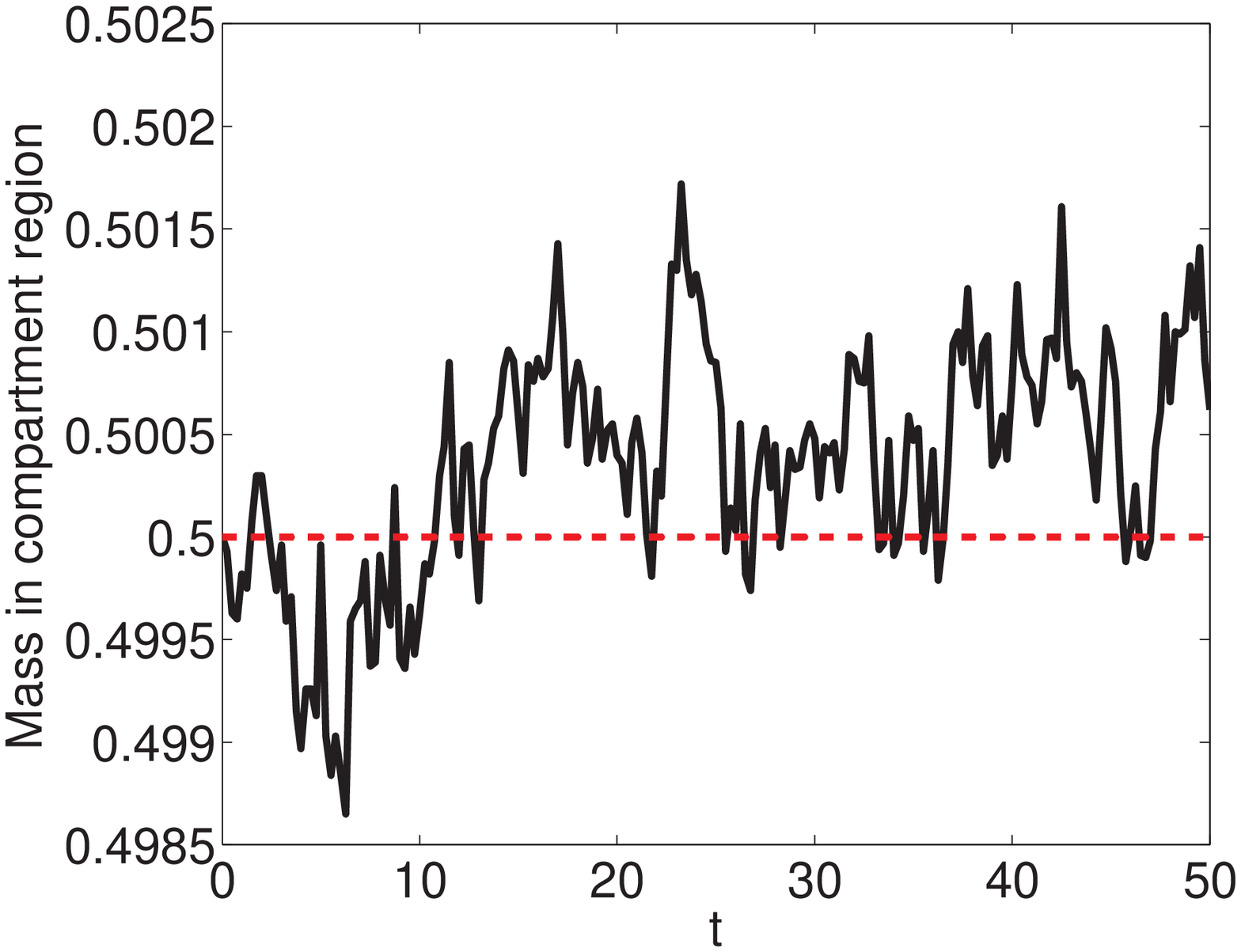}
\label{figure:uniform_compartment_mass_and_error_alg2}
}
\end{center}
\caption{Error plots for the uniform particle distribution maintenance example. In panel \subref{figure:uniform_PDE_mass_and_error_alg2} we display the normalised (i.e. divided by the total number of particles in the domain) expected value of the total mass in $\Omega_P$ given by the analytic solution (red, dashed line, $N_{DP}(t)/(N_{DP}+N_{DC}(t))$) and the normalised average mass in $\Omega_P$ according to our hybrid Algorithm 2 (black line, $N_{HP}(t)/(N_{HP}+N_{HC}(t))$). In panel \ref{figure:uniform_compartment_mass_and_error_alg2} we display the same comparison but for the mass in the compartment-based regime (i.e. $N_{DC}(t)/(N_{DP}+N_{DC}(t))$ (red, dashed line) and  $N_{HC}(t)/(N_{HP}+N_{HC}(t))$ (black line)). 
The masses generated by our hybrid algorithm are averaged over 100 repeats as in previous density comparison figures. No systematic bias is discernible even for simulations run over longer time periods and with more particles. }
\label{figure:uniform_error_analysis}
\end{figure}

The results for test problem 1 are shown in Figure \ref{figure:uniform_error_analysis}. In simulations generated by our hybrid Algorithm 2, the mass in the PDE-based regime is very similar to the expected mass given by the uniform solution of the diffusion equation. The same is true of the mass in the compartment-based regimes. Results are quantitatively indistinguishable (excepting for expected stochastic variation) for Algorithm 1, although for brevity we do not explicitly show the corresponding figures.

\begin{figure}[h!!!!!!!!!!!!!!!!!]
\begin{center}
\subfigure[]{
\includegraphics[width=0.45\columnwidth]{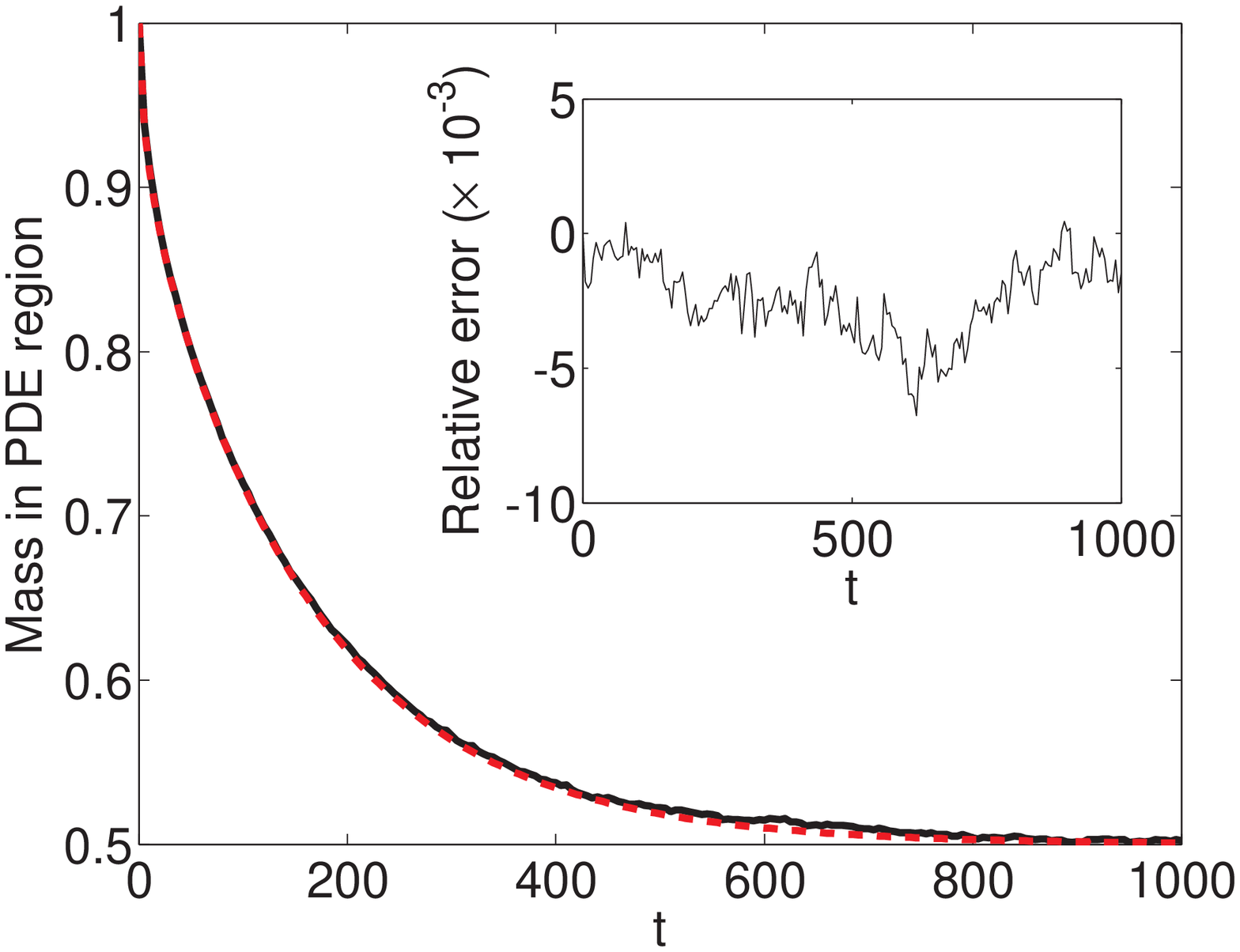}
\label{figure:step_fn_IC1_PDE_mass_and_error_alg2}
}
\subfigure[]{
\includegraphics[width=0.45\columnwidth]{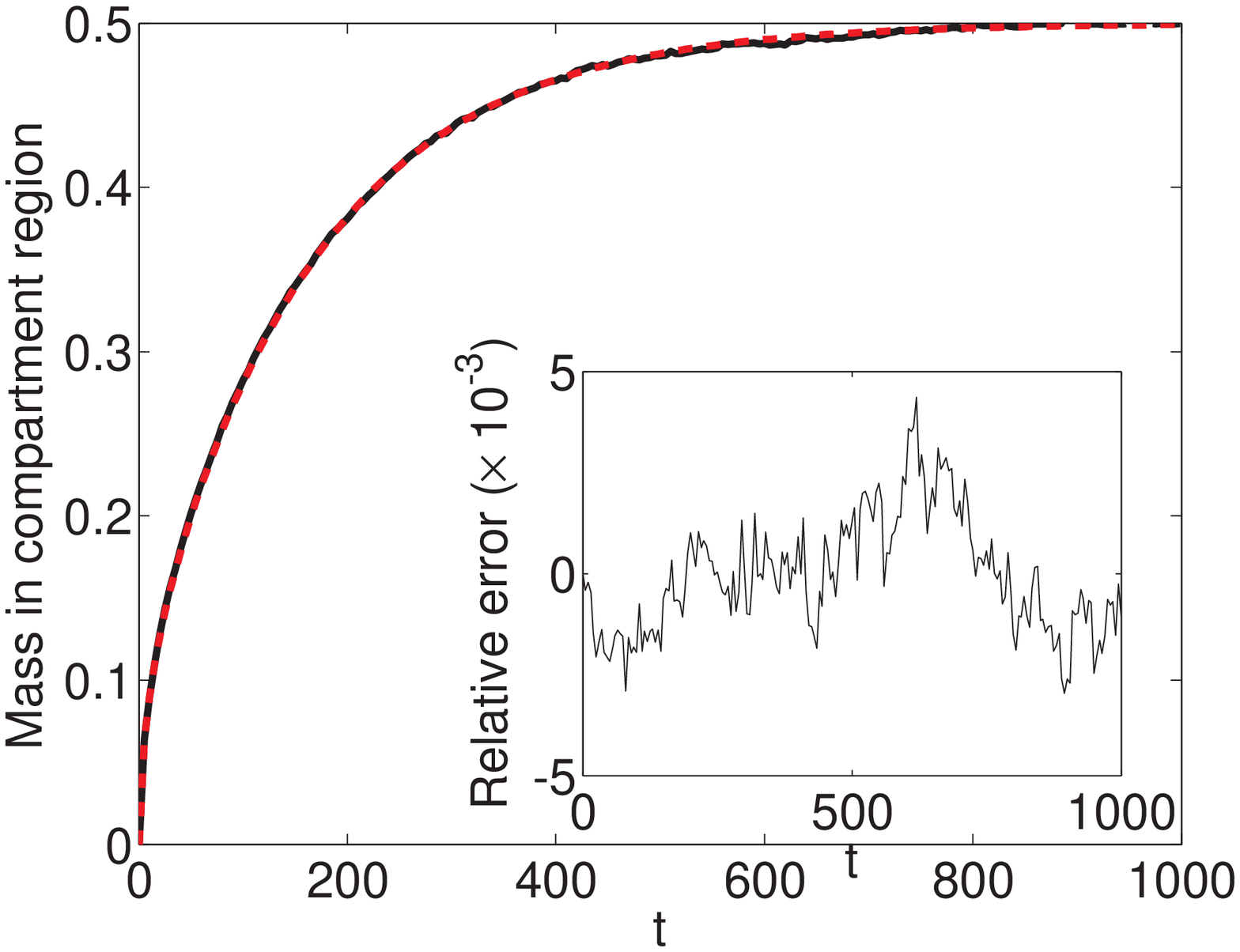}
\label{figure:step_fn_IC1_compartment_mass_and_error_alg2}
}
\end{center}
\caption{Error plots for test problem 2, in which all particles are initially in $\Omega_P$. In \subref{figure:step_fn_IC1_PDE_mass_and_error_alg2} we display the normalised expected value of the total mass in $\Omega_P$ given by the analytic solution (red, dashed line, $N_{DP}/(N_{DP}+N_{DC}$) and the normalised average mass in $\Omega_P$ (black line, $N_{HP}/(N_{DP}+N_{DC}$) according to our second hybrid algorithm. In \subref{figure:step_fn_IC1_compartment_mass_and_error_alg2} we display the same comparison but for the compartment-based regime. In each panel, the inset displays the evolution of the relative error (i.e. $(N_{HP}-N_{DP})/N_{DP}$ or $(N_{HC}-N_{DC})/N_{DC}$, respectively) throughout the simulation. 
Again, the masses generated by our hybrid algorithm are averaged over 100 repeats.}
\label{figure:step_function_IC_1_error_analysis}
\end{figure}

In Figure \ref{figure:step_function_IC_1_error_analysis} we display similar comparisons for test problem 2, in which the particles are initialised uniformly in $\Omega_P$ only. As the simulation evolves we see that the expected mass predicted by the hybrid algorithm matches closely the expected mass given by the analytic solution in both $\Omega_P$ and $\Omega_C$. Further, the insets demonstrate that the relative error between the masses generated by the analytic solution and the hybrid model are low and have no systematic bias, but instead fluctuate about their expected values. Similar results hold for the initial condition in which the particles are initialised uniformly in $\Omega_C$ only (see Figure 2
of the SM). As in Figure \ref{figure:uniform_error_analysis}, since the results of the simulations of our two hybrid algorithms are quantitatively similar, we present only the simulation results from our second hybrid algorithm.

\begin{figure}[h!!!!!!!!!!!!!!!!!]
\begin{center}
\subfigure[]{
\includegraphics[width=0.45\columnwidth]{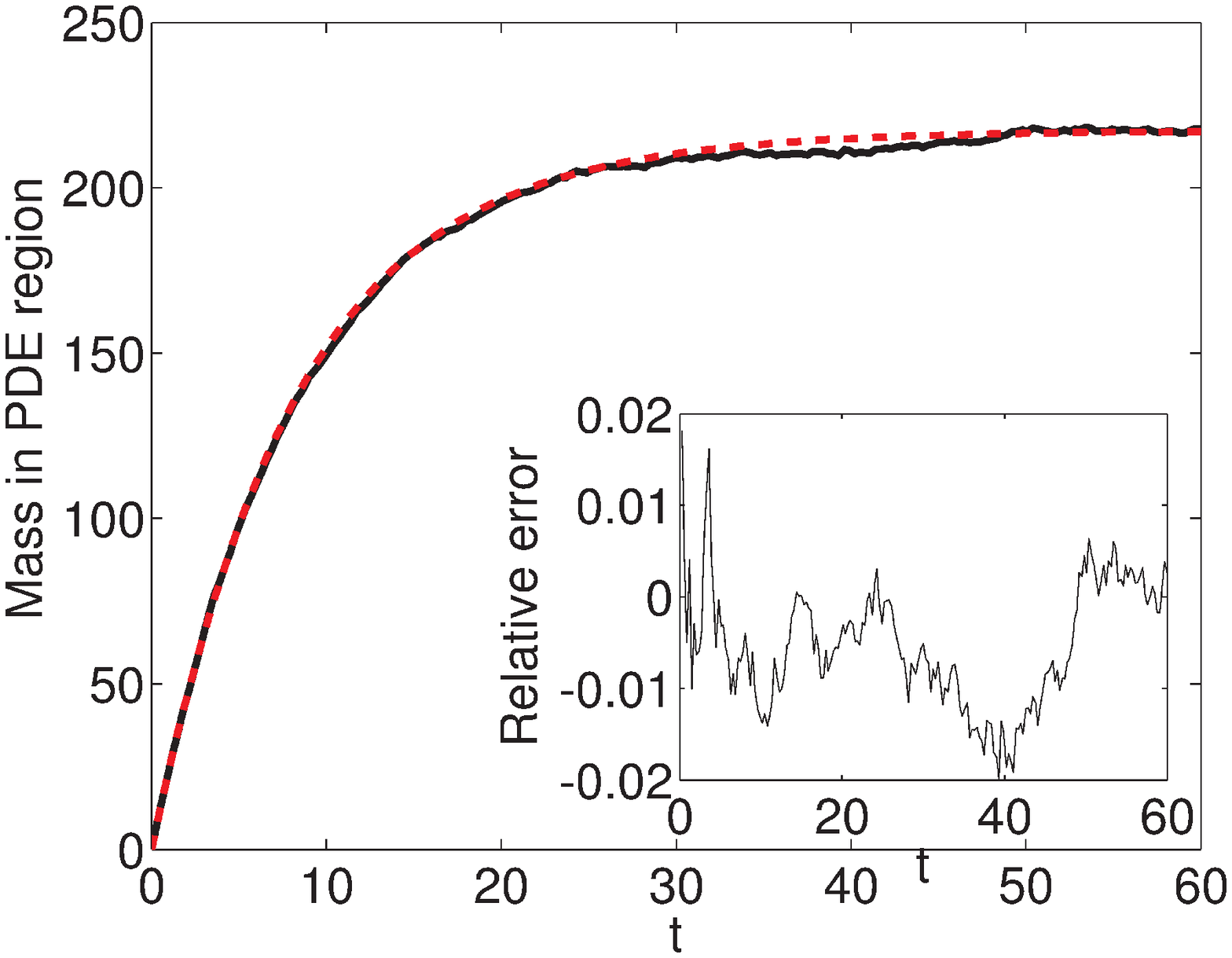}
\label{figure:morphogen_PDE_mass_and_error_alg1}	
}
\subfigure[]{
\includegraphics[width=0.45\columnwidth]{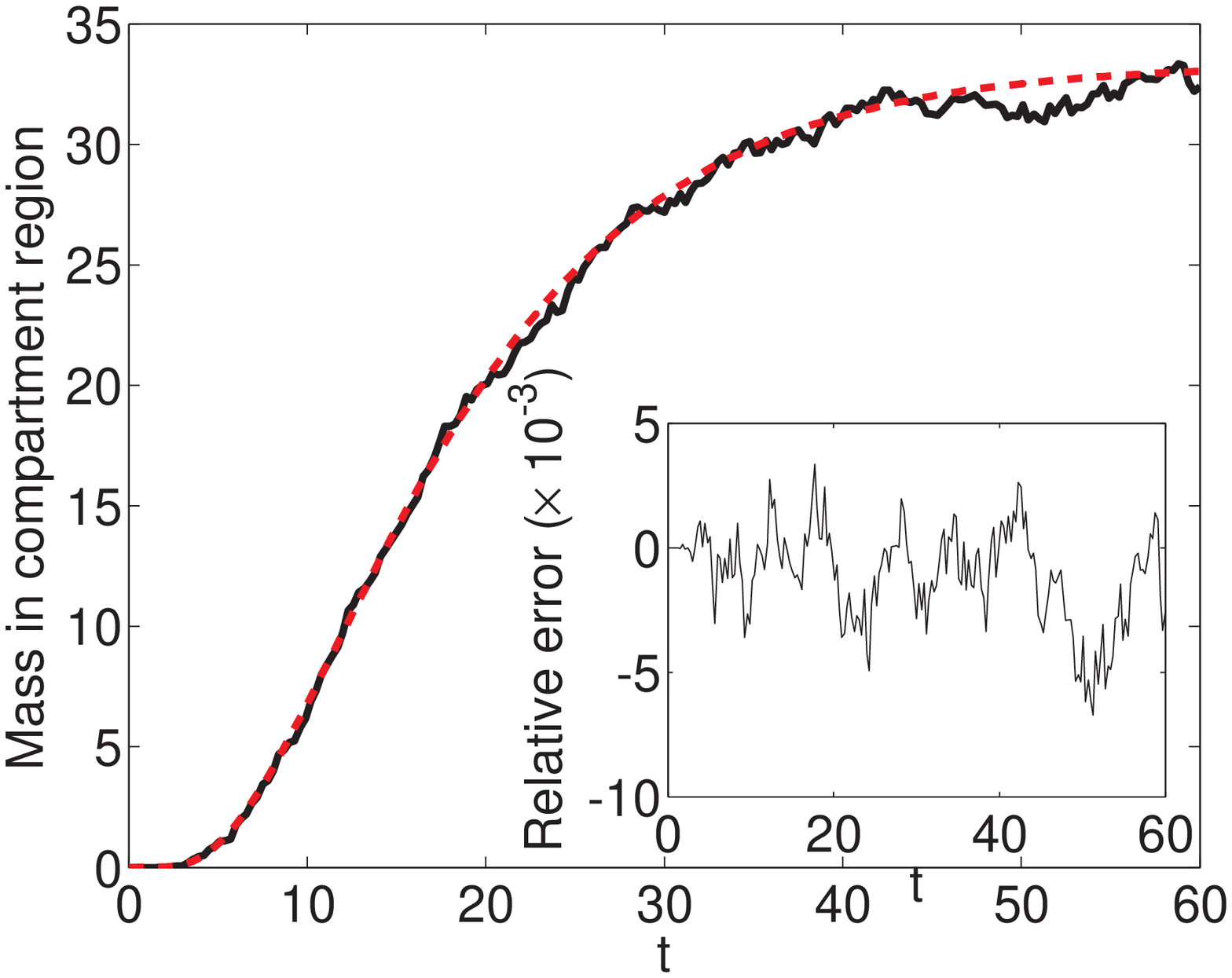}
\label{figure:morphogen_compartment_mass_and_error_alg1}
}
\subfigure[]{
\includegraphics[width=0.45\columnwidth]{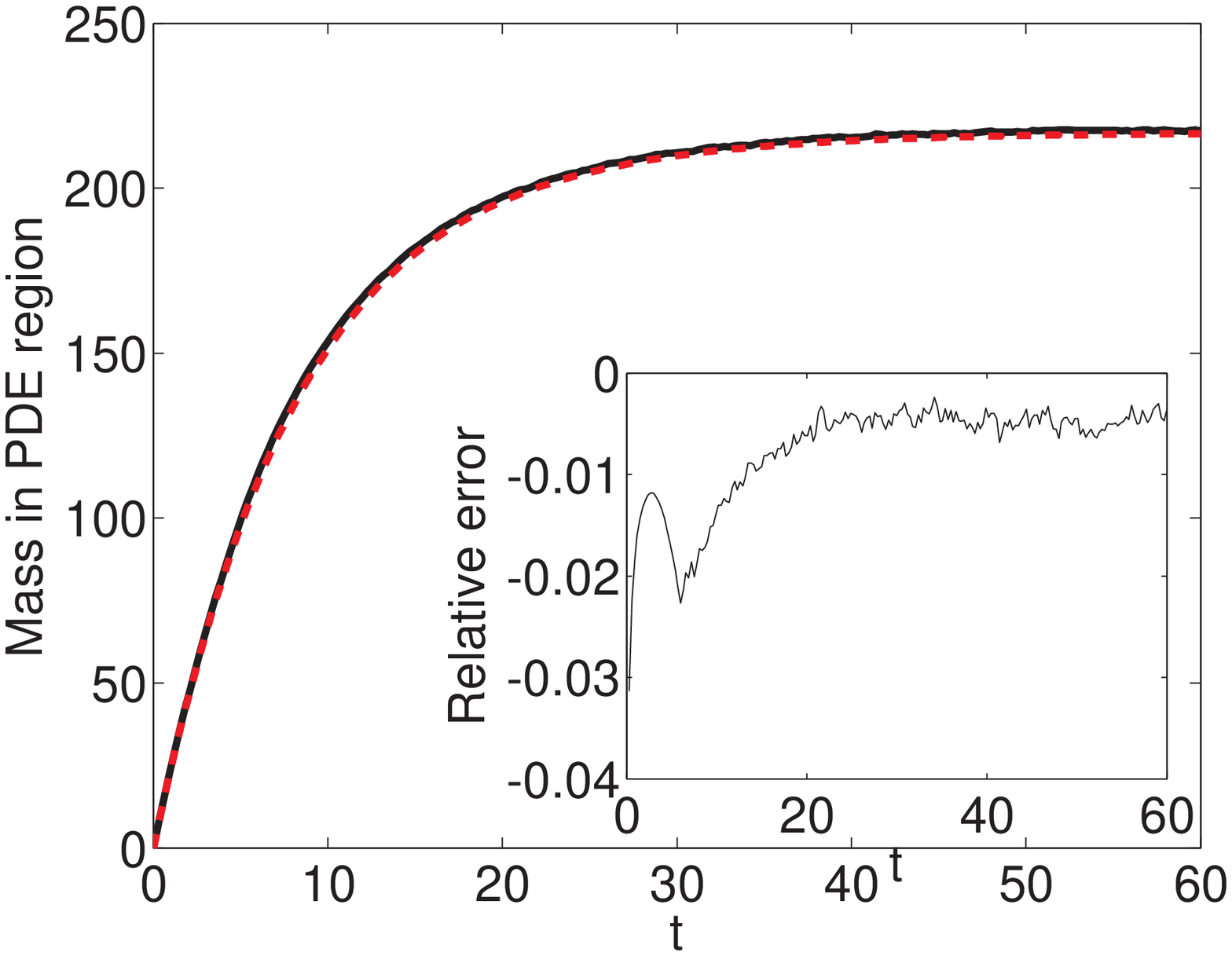}
\label{figure:morphogen_PDE_mass_and_error_alg2}
}
\subfigure[]{
\includegraphics[width=0.45\columnwidth]{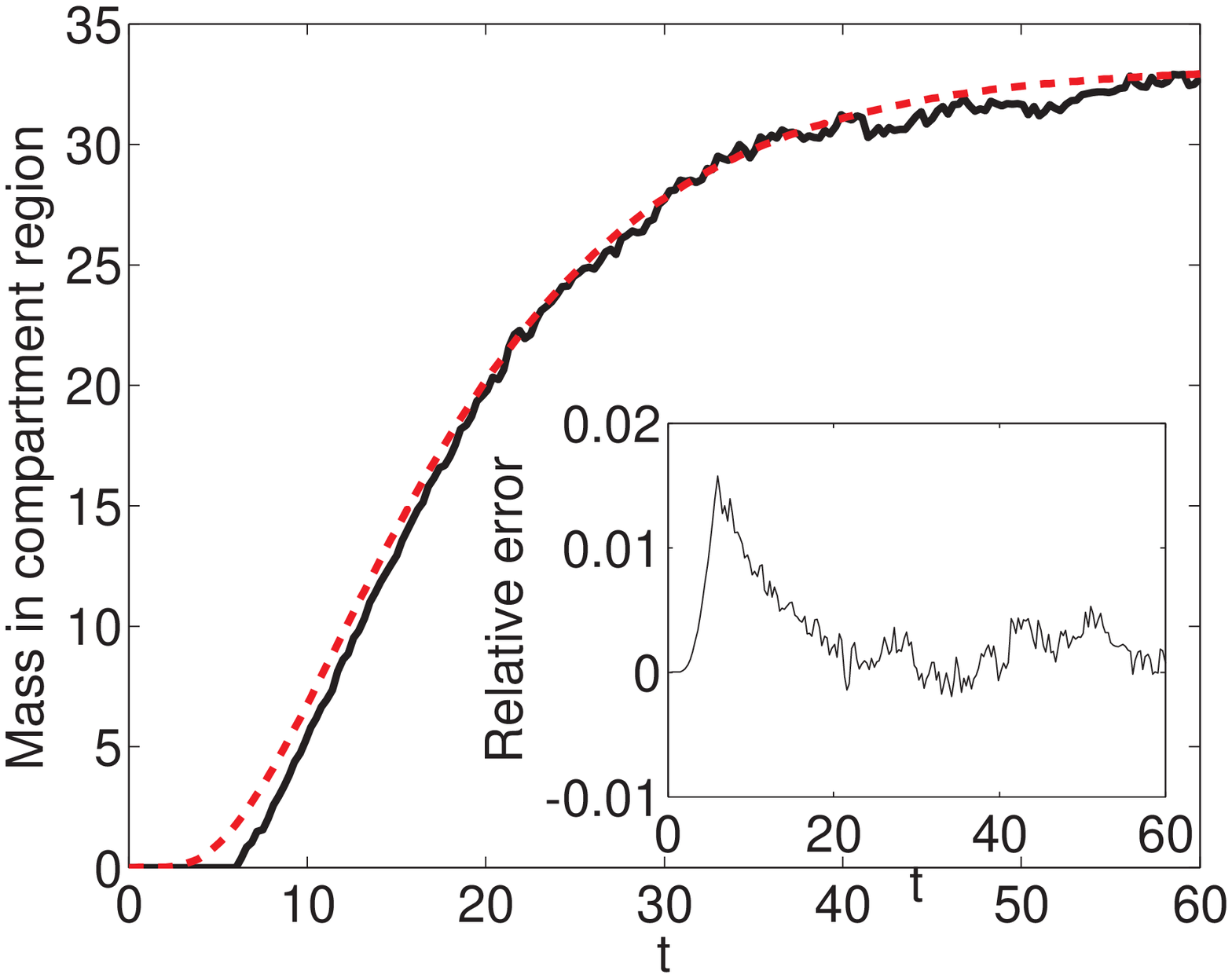}
\label{figure:morphogen_compartment_mass_and_error_alg2}
}
\end{center}
\caption{Error plots for the morphogen gradient formation example. Figure descriptions are as in Figure \ref{figure:step_function_IC_1_error_analysis}}
\label{figure:morphogen_gradient_error_analysis}
\end{figure}

Finally in Figure \ref{figure:morphogen_gradient_error_analysis} we display similar comparisons between the mass in each of the regimes for \textit{both} of the hybrid algorithms we have developed when simulating the formation of a morphogen gradient (test problem 3). Again, a comparison of the results from the hybrid algorithm and the analytic solution shows good agreement, especially for long times. However, there is a clear disparity between the mass in $\Omega_C$ for Algorithm 2 and the expected mass in that region (see Figure \ref{figure:morphogen_gradient_error_analysis} \subref{figure:morphogen_compartment_mass_and_error_alg2}). In particular the hybrid algorithm under-represents the expected mass in the compartment-based regime for early times. Although it is not clearly visible in Figure \ref{figure:morphogen_gradient_error_analysis} \subref{figure:morphogen_PDE_mass_and_error_alg2} due to the scale of the axes, there is a corresponding over-representation of the expected mass in $\Omega_P$ at early 
times for Algorithm 2. This is not the case for Algorithm 1 (see Figures \ref{figure:morphogen_gradient_error_analysis} \subref{figure:morphogen_PDE_mass_and_error_alg1} and \subref{figure:morphogen_compartment_mass_and_error_alg1}). 

The reason for this disparity is a technical one. Recall that in Algorithm 2, in order to avoid negative solution values, a particle is not allowed to be removed from the pseudo-compartment, $C_{-1}$, until the PDE solution across that compartment has value at least $1/h$. This avoids the possibility that the removal of a pseudo particle may send the value of the PDE solution in $C_{-1}$ negative. As the domain is initially empty and fills from the PDE region there is a significant period of time in which mass should be flowing over the interface, but is restricted from doing so by an artificial barrier in our approximate Algorithm 2. The flow of particles into $\Omega_C$ is, therefore, retarded in comparison to the analytic solution for which, of course, no such barrier exists. In Algorithm 1, however, we see no such problem (\textit{cf} Figure \ref{figure:morphogen_gradient_error_analysis} \subref{figure:morphogen_compartment_mass_and_error_alg1}), because mass which is moved across the interface is 
removed from the whole of $\Omega_P$, rather than just the pseudo-compartment.
This justifies our assertion that Algorithm 2 is suitable for simulations in which the number of particles at the interface are high, whereas Algorithm 1 is suitable for situations in which there may be low copy numbers at the interface.
We explore the ramifications of these results further in the discussion.

\begin{figure}[h!!!!!!!!!!!!!!!!!]
\begin{center}
\subfigure[]{
\includegraphics[width=0.445\columnwidth]{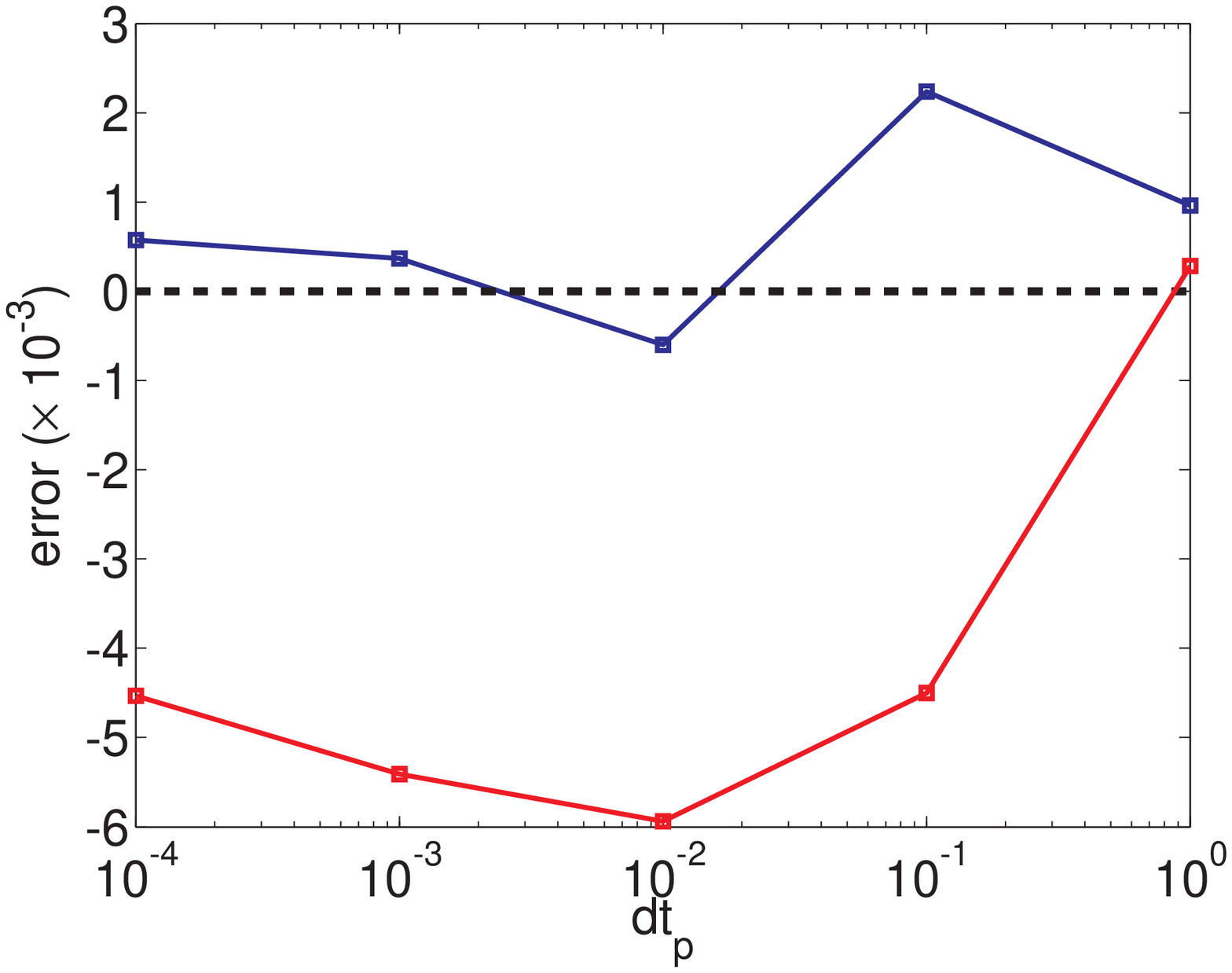}
\label{figure:error_with_dt_p}
}
\subfigure[]{
\includegraphics[width=0.45\columnwidth]{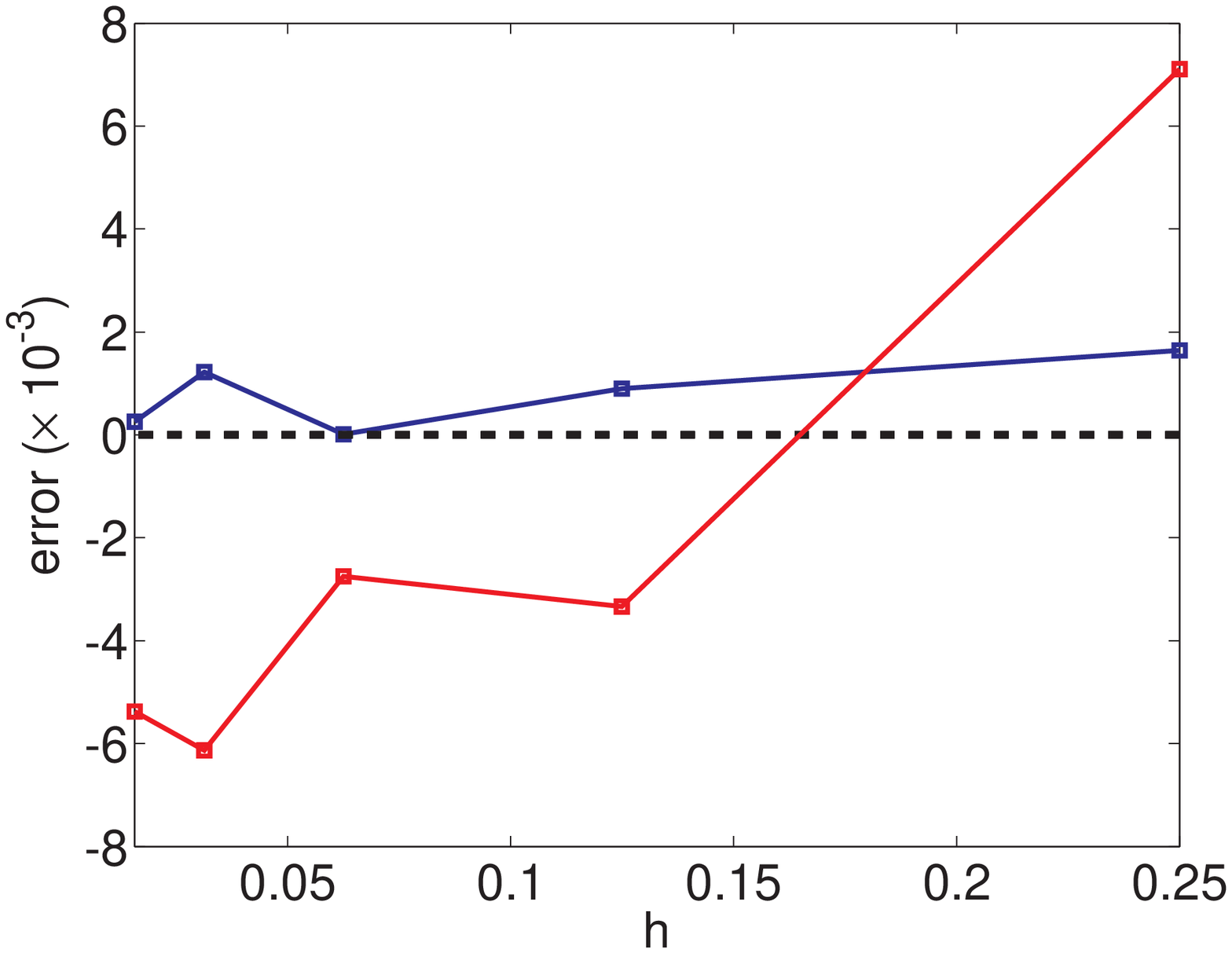}
\label{figure:error_with_h}
}
\end{center}
\caption{Error in compartment-based particle numbers varies with \subref{figure:error_with_dt_p} PDE step size, $dt_p$, and \subref{figure:error_with_h}  compartment size, $h$, for both test problem 1 (blue) and test problem 2, in which all the particles are uniformly distributed in $\Omega_P$ (red). For \subref{figure:error_with_dt_p} simulation parameters are as given previously with $dt_p$ varying in steps of powers of ten from $10^{-4}$ to $1$ (note the log-scale on the horizontal axis.). For convenience in \subref{figure:error_with_h} we chose $\Delta x=1/256$ and varied $h$ in powers of two from $1/64$ to $1/4$. In test problem 1 final simulation time was 50 (as in Figure \ref{figure:uniform_error_analysis}), whereas in test problem 2 the final simulation time was 250.}
\label{figure:error_with_model_parameters}
\end{figure}

In order to determine the robustness of our coupling mechanism we performed further analysis of the error for test problems 1 and 2 (`maintenance of a uniform steady' state and `particles flowing from $\Omega_P$ to $\Omega_C$', respectively). Specifically we investigated the emergence of error associated with the discretisation parameters $dt_p$ (the time discretisation associated with PDE updates) and $h$ (the compartment size in $\Omega_C$) (see Figures \ref{figure:error_with_model_parameters} \subref{figure:error_with_dt_p} and \subref{figure:error_with_h}, respectively). The time-averaged relative error in the expected number of particles in $\Omega_C$ simulated using our hybrid simulation was determined by comparing to the same quantity predicted by a continuous mean-field analytical PDE solution to the respective test problems. In order to reduce the effect of stochastic fluctuation on the error measurement we averaged the particle numbers over 100 repeat simulations each each with $N=1000$ particles 
for each of the test problems. Due to the computational intensity of these algorithms with such larger particle numbers and the wide variety of simulation parameters we conducted our error analysis using the algorithm two. We expect the error behaviour to be the same for algorithm 1 since the source of error we have identified (see below) should affect both algorithms in a similar manner.

The results of varying both $h$ and $dt_p$ separately are shown in Figure \ref{figure:error_with_model_parameters}. On the vertical axis, error is defined as $\left\langle(N_{DC} - N_{HC})/N_{DC}\right\rangle$ where, as before, $N_{DC}$ is the known expected number of particles in $\Omega_C$ and $N_{HC}$ is the averaged simulated number of particles in the same region. The angle brackets denote the time average of the relative error recorded at unit time intervals throughout the simulation. A positive error indicates a net bias in the simulations for particles towards $\Omega_P$ whereas a negative error indicates a net bias for particles towards $\Omega_C$.

Test problem 1 contains smaller sources of error than test problem 2. These errors have a consistent order of magnitude as that of expected stochastic fluctuations as a result of finite copy number. The difference in the magnitude of the error between test problem 1 and test problem 2 can be attributed to the way in which mass transfer is balanced or unbalanced. For test problem 1, net diffusive transfer of mass into the pseudo-compartment from the PDE is zero as is the net transfer of particles between the pseudo-compartment and $\Omega_C$. In test problem 2 there is a net transfer of mass into the pseudo-compartment from the PDE regime and a net transfer out to the compartment-based regime. The systemic sources of error that we have identified in our method result from a non-zero net particle flux over the interface. The pseudo-compartment method results in these errors being very small. 

Whilst still very small, a comparatively larger error associated with the hybrid modelling approach is observed in cases of non-zero net flux (in which the well-mixed assumption does not strictly hold within the pseudo-compartment). Increasing both $h$ and $dt_p$ results in an increase in relative error in test problem 2 (see Figure \ref{figure:error_with_model_parameters}). The reason for this increase is not directly related to the hybrid model but rather to the individual single-scale models used. Significant increases in the error for test problem 2 can be seen when $dt_p$ is increased beyond $O(dx_p^2/D)$. Similarly, an $O(h^2)$ error is expected for large compartment sizes, $h$. In both cases, the effective diffusion is reduced (in comparison to that expected by the equivalent continuous PDE) as a result of discretisation error in the PDE and compartment region, respectively. The reduction of effective diffusion leads to a reduction in the net diffusive flux over the interface regardless of the 
implementation of the interface itself. The increase in error with increasing $h$ and $dt_p$ is therefore, a manifestation of discretisation error in the simulation methods employed in $\Omega_C$ and $\Omega_P$, respectively. 

The sign of the relative error in test problem 2 is negative for low $h$ and low $dt_p$. This is a manifestation of a small increase in diffusive flux caused by our algorithm pseudo-compartment. The cause of this error is due to the breakdown of the assumption that mass in $C_{-1}$ is evenly distributed. Indeed, in test problem 2, mass flowing into $C_{-1}$ from the remainder of $\Omega_P$ is instantaneously considered to be evenly spread in $C_{-1}$ for the purposes of consideration of jumping into $\Omega_C$. Instead, this mass lies a small distance away from the expected position at the centre of the node on the side which is furthest from the interface. This erroneous evenly-distributed assumption results in a slightly increased net flux of particles over the interface that would otherwise be expected. The reason that this source of error is not prevalent in test problem 1 is because there is no \textit{net} flux of particles flowing into or out of $C_{-1}$ that would contribute to such an error. The 
effect of this error is always to create a net increased bias over the interface in the direction of the net flow (an overall negative relative error in the case of test problem 2). It is important to notice that this error is common at the interface of hybrid models of this type. The inclusion of the pseudo-compartment rather than a clean interface forces particles which would contribute to this error to have a decreased number of transfer events associated with them and therefore compound a much smaller error (see for example, the two regime method  \cite{flegg2012trm}). 

Fine details of the error and its dependence on the small model parameters $\Delta x$, $dt_p$ and $h$ is complex and difficult to determine due to other algorithmic factors (for example, particles need to wait between PDE times steps to be considered for $C_{-1}\rightarrow C_{1}$ when they are near the edge of $C_{-1}$). It is also unclear if the reflecting boundary at $I$ plays any role as a source of error for particular discretisation parameters. Some of these issues may be responsible for complex error behaviour in Figure \ref{figure:error_with_model_parameters} for small $h$ and small $dt_p$. On the other hand, details of fluctuation of the error for small $h$ and small $dt_p$ are on the order of fluctuations that are expected as a result of finite copy number and may not be systemic in nature.

 Importantly, for small $h$ and $dt_p$, whilst there is a bias for particles to flow over the interface in the direction of net flow, this bias is still very small. The authors recommend the use parameters $\Delta x$, $dt_p$ and $h$ that individually minimize the discretisation error of the PDE and compartment-based methods used. The interface should create an bias approximately proportional to $h$ in the direction of net flow but this bias is significantly reduced due to the dual nature of pseudo-compartment. For completeness, the authors recommend that $dx_p << h$ to maintain the approximation of a PDE in $\Omega_P$ although it would appear that the method is robust when this approximation is relaxed. Indeed, if $dx_p = h$ one would expect no interface-related error at all since transfer into the pseudo-compartment from the ``PDE'' is concentrated at the node at the centre of the pseudo-compartment. In this case though, the PDE becomes more closely described as a diffusion master equation on a lattice.

\section{Discussion}

We have introduced two algorithms which allow a coupling between a continuum PDE description of particle density in one region of the domain and a discrete stochastic compartment-based description in another. Using a \textit{pseudo-compartment} in the PDE regime, which has properties of both descriptions, we were able to couple the two simulation methodologies together in such a way that the expected flux across the interface is maintained.
We evidenced this though a series of examples, demonstrating a good correspondence between the expected mass in each of the modelling regions (averaged over several repeats of the hybrid simulation algorithms) and the mass in those regions determined by the corresponding fully-deterministic mean-field model. We also analysed the errors in the coupling method over a wide range of model parameter values and found them to be small (always less than 1\%) even at the extremes of the parameter regimes in which the single-scale models we employ on either side of the interface are not expected to work well. 

Our method relies on a knowledge of the deterministic PDE description to which the mean of the compartment-based simulations corresponds. Whilst this is easy to obtain in the examples we have studied, when higher order reactions are introduced to the stochastic model there is no longer an exact closed-form representation of the expected particle density as a PDE in terms of the first moment alone. In order to obtain a self-contained equation we must make moment closure assumptions, which means that the PDE description will no longer correspond exactly to the mean behaviour of the compartment-based system. However, in situations where particle numbers are sufficiently large the closed PDE description will often be a close match to the expected densities of the compartment-based model and, as such, we expect that our algorithms will still be applicable. Furthermore, large particle numbers are precisely the regime in which we want to make use of the PDE description, further improving the applicability of our 
algorithms.

It is precisely this consideration which will provide motivation for our next work on this subject to develop a hybrid method with an adaptive interface. In order to make the hybrid algorithm as efficient as possible, the interface should be able to move so that only regions with sufficiently low concentrations are represented by the compartment-based model and regions with high concentration are consistently represented with the PDE-based model. This would ultimately allow for Algorithm 2 to be used in all cases for efficiency reasons. A good set of rules that could be used to move an interface was explored in \citep{robinson2014atr}. This will increase the computational efficiency of the algorithm, since we will no longer be simulating the individual particle movements in the discrete model when the continuum description can be tolerated. There is also scope to interface the algorithms we have developed in this manuscript with algorithms designed to couple compartment-based models and molecular-based 
models \citep{flegg2013cmc,flegg2012trm} in order to build a ``three-regime method''. Another important extension of this work is to consider compartment-based models on irregular lattices which will allow us to capture more complex and realistic domain geometries.

For the sake of clarity in this manuscript, we presented the algorithms themselves and the illustrative examples in a single dimension, however, the algorithm is easily adapted for higher dimensional problems with (hyper)-planar interfaces. We therefore expect our algorithms to be highly applicable to stochastic simulations of systems in which molecular populations are high in some regions and low in others, meaning that neither a fully-deterministic-continuum nor a fully-stochastic-individual-based model is appropriate. Our algorithms provide significant increases in speed in comparison to fully stochastic models and more accurate and representative modelling in comparison to fully-deterministic models.

\bibliography{hybrid}
\bibliographystyle{plainnat}

\end{document}